\documentclass[twocolumn,showpacs,preprintnumbers,amsmath,amssymb]{revtex4-1}

\usepackage{graphicx} 
\usepackage{dcolumn} 
\usepackage{bm}
\usepackage{amsmath}
\usepackage{amssymb}
\usepackage{braket}

\newcommand{\up}{\uparrow}
\newcommand{\down}{\downarrow}

\newcommand{\ep}{\varepsilon}
\def\lesssim{\ \raise.3ex\hbox{$<$}\kern-0.8em\lower.7ex\hbox{$\sim$}\ }
\def\gesim{\ \raise.3ex\hbox{$>$}\kern-0.8em\lower.7ex\hbox{$\sim$}\ }
\begin{document}
\title{Non-equilibrium strong-coupling theory for a driven-dissipative ultracold Fermi gas in the BCS-BEC crossover region}
\author{Taira Kawamura\email{tairakawa@keio.jp}$^1$, Ryo Hanai$^{2,3}$, Daichi Kagamihara$^1$, Daisuke Inotani$^4$, and Yoji Ohashi$^1$}
\affiliation{$^1$ Department of Physics, Keio University, 3-14-1 Hiyoshi, Kohoku-ku, Yokohama 223-8522, Japan}
\affiliation{$^2$ Department of Physics, Osaka University, Toyonaka 560-0043, Japan}
\affiliation{$^3$ James Franck Institute and Department of Physics, University of Chicago, Chicago, Illinois 60637, USA}
\affiliation{$^4$ Research and Education Center for Natural Sciences, Keio University, 4-1-1 Hiyoshi, Yokohama, Kanagawa 223-8521, Japan}

\date{\today}

\begin{abstract}
We theoretically investigate strong-coupling properties of an ultracold Fermi gas in the BCS-BEC crossover regime in the non-equilibrium steady state, being coupled with two fermion baths. By developing a non-equilibrium strong-coupling theory based on the combined $T$-matrix approximation with the Keldysh Green's function technique, we show that the chemical potential bias applied by the two baths gives rise to the anomalous enhancement of Fulde-Ferrell-Larkin-Ovchinnikov (FFLO) type pairing fluctuations (although the system has no spin imbalance), resulting in the re-entrant behavior of the non-equilibrium superfluid phase transition in the BCS-unitary regime. These pairing fluctuations are also found to anomalously enhance the pseudogap phenomenon. Since various non-equilibrium phenomena have recently been measured in ultracold Fermi gases, our non-equilibrium strong-coupling theory would be useful to catch up this experimental development in this research field.
\end{abstract}
\maketitle
\par
\section{Introduction}
\par
Since the realization of the superfluid phase transition in $^{40}{\rm K}$ \cite{PhysRevLett.92.040403} and $^6{\rm Li}$ \cite{PhysRevLett.92.120403,PhysRevLett.92.150402,PhysRevLett.92.203201} Fermi gases, cold Fermi gas physics has dramatically progressed \cite{Levin2005,Rev1,Rev2}. An advantage of this system is the high tunability of various physical parameters. In particular, a tunable pairing interaction associated with a Feshbach resonance  \cite{RevModPhys.82.1225} has enabled us to systematically study Fermi superfluids from the weak-coupling BCS (Bardeen-Cooper-Schrieffer) regime to the strong-coupling BEC (Bose-Einstein condensation) limit \cite{Leggett,NSR,Randeria1995,Engelbrecht,Strinati,Griffin}. The intermediate coupling regime, which is referred to as the BCS-BEC crossover region in the literature, has particularly attracted much attention \cite{Levin2005,Rev1,Rev2,Zwerger}, because strong pairing fluctuations dominate over system properties there.
\par
So far, ultracold Fermi gases have mainly been studied in the {\it thermal equilibrium} case, because the usual experimental situation of a trapped Fermi gas is considered to be well isolated from the environment, and also be in the (quasi-)equilibrium state. However, a strong interest on {\it non-equilibrium} properties of this strongly interacting quantum many-body system has recently emerged, fueled by the experimental realization of tunable dissipation \cite{Chong2018,Tomita2017,PhysRevLett.116.235302,PhysRevLett.115.050601,PhysRevLett.110.035302}. 
\par
\begin{figure}[b]
\centering
\includegraphics[width=75mm]{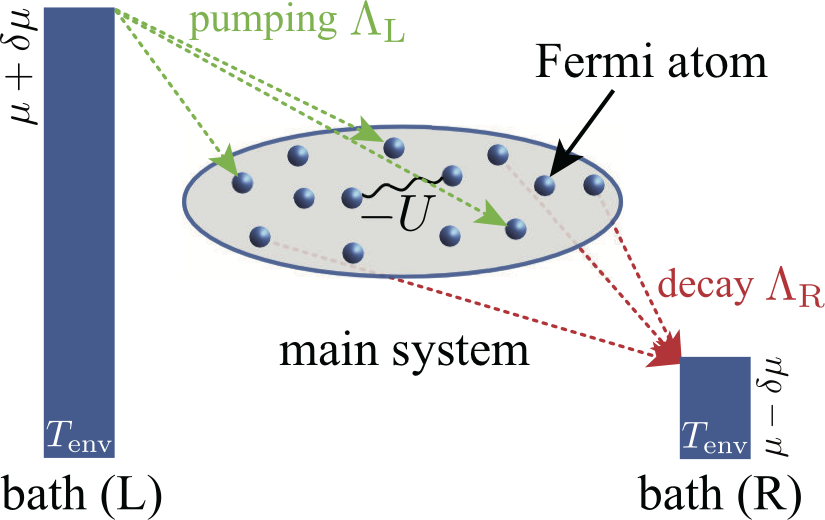}
\caption{Model non-equilibrium driven-dissipative Fermi gas with a tunable $s$-wave pairing interaction $-U~(<0)$ associated with a Feshbach resonance. The non-equilibrium (main) system is coupled with (1) a pumping bath (L) with the chemical potential $\mu_{\rm L}=\mu+\delta\mu$ and a coupling strength $\Lambda_{\rm L}$ (which supplies Fermi atoms to the system), as well as (2) a decay bath (R) with $\mu_{\rm R}=\mu-\delta\mu$ and $\Lambda_{\rm R}$ (which absorbs Fermi atoms from the system). The baths are assumed to be free Fermi gases in the thermal equilibrium state at the temperature $T_{\rm env}$.}
\label{fig1} 
\end{figure}
\par
The study of quantum many-body correlation effects in the presence of dissipation is currently a rapidly evolving field \cite{Houck2012,Carusotto2013,Sanvitto2016,Daley2014}, both experimentally and theoretically \cite{Sieberer2016} in various platforms, such as superconducting circuits \cite{Houck2012}, exciton polaritons \cite{Carusotto2013,Sanvitto2016}, strongly correlated photons \cite{Jia2018}, and trapped ions \cite{Barreiro2011}. To highlight, a Mott insulator of photons in a superconducting circuit \cite{Ma2019} was recently realized by a careful dissipation engineering, and an exotic quantum state of matter \cite{Nakagawa2019}  and anomalous critical phenomena \cite{Sieberer2013,Tripathi2016, Ashida2016, Hanai2019} are predicted to arise from the interplay between the many-body correlation effects and dissipations.  Regarding the high tunability of the pairing interaction and dissipation strength, a cold atomic Fermi gas seems to be a promising playground to largely broaden the understandings of such nonequilibrium many-body effects in a systematic way.  
\par
Motivated by the above-mentioned possibilities, in this paper, we theoretically investigate a strongly interacting ultracold Fermi gas in a driven-dissipative steady state, by extending a strong-coupling $T$-matrix approximation (TMA) \cite{PhysRevB.66.024510, PhysRevA.80.033613}, developed in the thermal equilibrium Fermi gas, to the non-equilibrium case by utilizing the Keldysh Green's function technique \cite{Rammer2007,Zagoskin,Stefanucci}.  Among various driven-dissipative non-equilibrium situations, in this paper, we pick up an open Fermi gas which is coupled with two fermion baths (pumping bath (L) and decay bath (R) in Fig. \ref{fig1}). Such a situation has recently been realized in a coupled optical lattice with a thermal atomic reservoir in a magneto-optical trap \cite{Chong2018}. A similar situation may also be expected by extending the recent transport experiment on a $^6$Li Fermi gas in a two-terminal configuration \cite{Brantut2012,Krinner2015,Husmann2015,Krinner2017}. Thus, by combining these experimental techniques \cite{Chong2018,Brantut2012,Krinner2015,Husmann2015,Krinner2017} with a Feshbach-induced tunable pairing interaction, the realization of BCS-BEC crossover out of equilibrium is promising. 
\par
\begin{figure}[b]
\centering
\includegraphics[width=85mm]{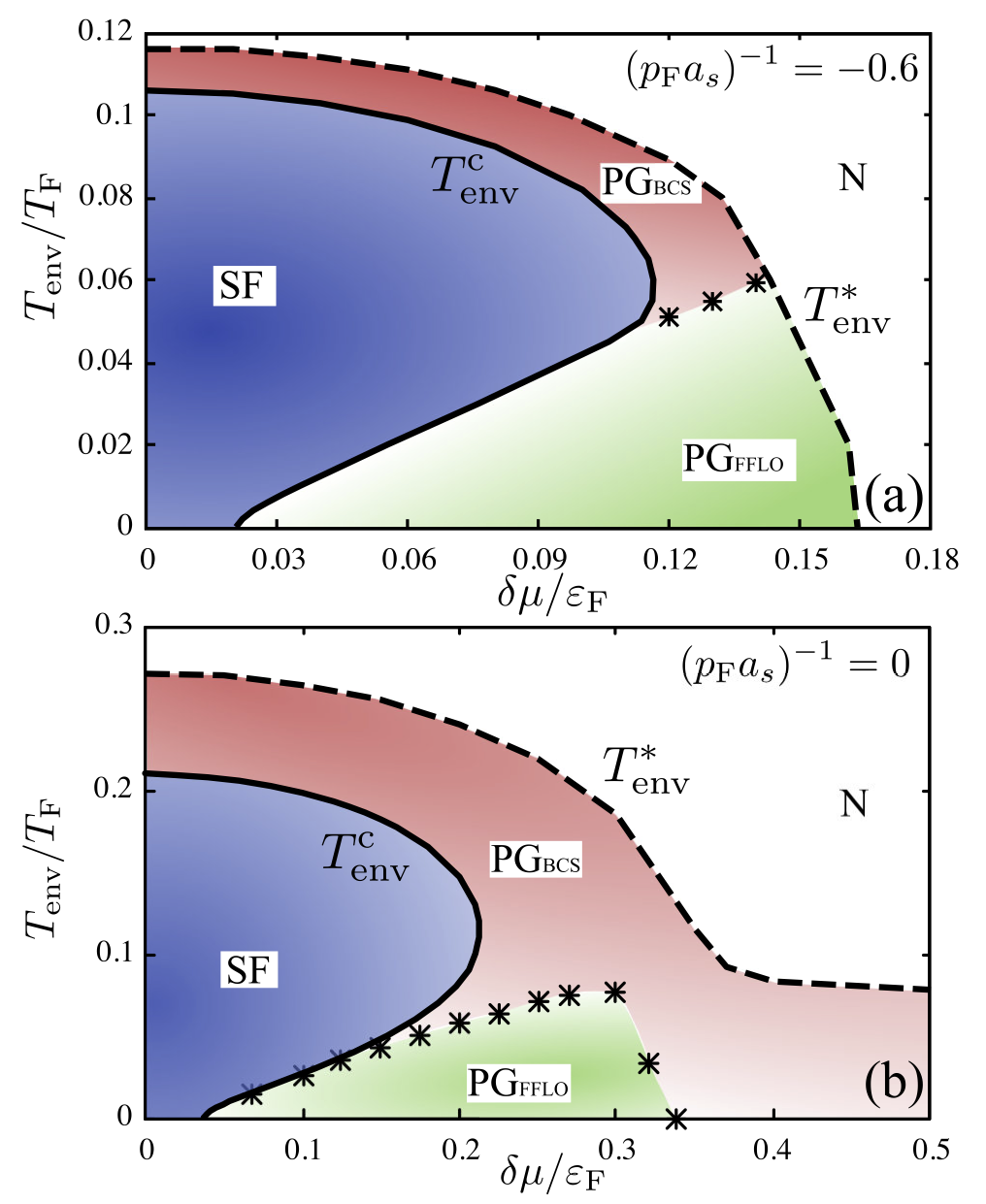} 
\caption{
Phase diagram of a non-equilibrium two-component Fermi gas with pumping and decay in terms of the temperature of the environments $T_{\rm env}$ and the chemical potential bias $\delta\mu$ applied by the two baths. (See Fig. \ref{fig1}) (a) BCS regime $(p_{\rm F}a_s)^{-1}=-0.6$. (b) Unitary limit $(p_{\rm F}a_s)^{-1}=0$.  $T^{\rm c}_{\rm env}$ (solid line) is the superfluid phase transition temperature, and $T^*_{\rm env}$ (dashed line) is the pseudogap temperature. The regions labeled by ``SF'', ``N'', ``PG$_{\rm BCS}$'', and ``PG$_{\rm FFLO}$'' corresponds to the superfluid state, the normal state, pseudogap regime where BCS-type (zero center-of-mass momentum) pairing fluctuations are dominant \cite{Levin2005,Strinati,PhysRevB.66.024510,PhysRevA.80.033613}, and pseudogap regime where FFLO-type (finite center-of-mass momentum) pairing fluctuations are dominant, respectively. For the concrete criteria for determining these regimes, see Sec. III.B.}
\label{fig2} 
\end{figure}
\par
Our principal results are captured in Fig. \ref{fig2}, which shows the steady-state phase diagram of a non-equilibrium two-component Fermi gas with pumping and decay in the BCS and unitary regime. In these regimes, we find that the chemical potential bias $\delta\mu$ applied by the two baths gives rise to an anomalous enhancement of Fulde-Ferrell-Larkin-Ovchinnikov (FFLO) type pairing fluctuations, although the system has no spin imbalance. These anomalous FFLO-type fluctuations turn out to cause re-entrant behavior of the non-equilibrium superfluid phase transition. We also find that these fluctuations anomalously enhance the pseudogap phenomenon (PG$_{\rm FFLO}$ region in Fig. \ref{fig2}). The FFLO-type fluctuations and related phenomena disappear in the BEC regime.
\par
We briefly note that the non-equilibrium Fermi systems with a strong attractive interaction have also been discussed in other fields, such as an exciton gas \cite{Yoshioka2011} (bound electron-hole pairs), exciton-polariton condensate \cite{Kasprzak2006,Carusotto2013} (superpositions of excitons and micro-cavity photons), as well as neutron star cooling \cite{Yakovlev2005}. In particular, the extension of the mean-field theory to the driven-dissipative non-equilibrium case has already been done in the context of exciton and exciton-polariton physics \cite{PhysRevB.96.125206,PhysRevLett.96.230602,Yamaguchi2012,PhysRevB.97.245302}. In this sense, this work is an extension of these previous work to include pairing fluctuations beyond the mean-field level. 
\par
This paper is organized as follows. In Sec. II, we explain how to extend TMA in the thermal equilibrium state to the case of non-equilibrium steady state, by using the Keldysh Green's function technique. We show our results in Sec. III. Here, we examine the superfluid phase transition temperature, as well as single-particle excitations, in the non-equilibrium BCS-BEC crossover region. Throughout this paper, we set $\hbar=k_{\rm B}=1$, and the system volume $V$ is taken to be unity, for simplicity.  
\par
\section{Formulation}
\par
In this section, we explain our theoretical framework. In Sec. II.A, we present a model open Fermi gas which is coupled with a pumping and decay baths. As a useful tool to deal with both non-equilibrium and strong-coupling effects, we introduce the Keldysh Green's function theory in Sec. II.B. Using this, we explain how to extend TMA to the non-equilibrium steady state in Sec. II.C. In Sec. II.D, we derive the equation for the superfluid phase transition temperature within the framework of the non-equilibrium $T$-matrix approximation (NETMA). We also confirm that NETMA is reduced to TMA in the thermal equilibrium limit in Sec. II.E.
\par
\par
\subsection{Model non-equilibrium interacting Fermi gas}
\par
We consider a two-component Fermi gas, which is coupled with a pumping bath and a decay bath, as illustrated in Fig. \ref{fig1}. To model this system, we consider the Hamiltonian \cite{PhysRevB.96.125206},
\begin{equation}
H=H_{\rm sys}+H_{\rm env}+H_{\rm mix}. 
\label{Eq_H_tot}
\end{equation}
Here, 
\begin{align}
H_{\rm sys} &=\! \sum_{\bm{p},\sigma} \ep_{\bm{p}} a^\dagger_{\bm{p},\sigma} a_{\bm{p},\sigma} \nonumber\\
&\!-U\! \sum_{\bm{p},\bm{p}',\bm{q}}\! a^\dagger_{\bm{p}+\bm{q}/2,\up}  a^\dagger_{-\bm{p}+\bm{q}/2,\down} a_{-\bm{p}'+\bm{q}/2,\down} a_{\bm{p}'+\bm{q}/2,\up}
\label{Eq_H_sys}
\end{align}
describes a (non-equilibrium) Fermi gas, which we call the (main) system in this paper. $a_{\bm{p},\sigma}$ is an annihilation operator of a Fermi atom with pseudo-spin $\sigma=\up,\down$, describing two atomic hyperfine states. $\ep_{\bm{p}}={\bm p}^2/(2m)$ is the kinetic energy, where $m$ is an atomic mass. $-U~(<0)$ is a pairing interaction, which is assumed to be tunable by using a Feshbach resonance \cite{RevModPhys.82.1225}. As usual, we measure the interaction strength in terms of the $s$-wave scattering length $a_s$, which is related to the pairing interaction $-U$ as \cite{Randeria1995},
\begin{equation}
\frac{4\pi a_s}{m} = \frac{-U}{1-U\sum_{\bm{p}}^{p_\text{c}} \frac{1}{2\ep_{\bm{p}}}},
\end{equation}
where $p_\text{c}$ is a momentum cutoff. The weak-coupling BCS side and the strong-coupling BEC side are then characterized by $(p_{\rm F}a_s)^{-1} \lesssim 0$ and $(p_{\rm F}a_s)^{-1} \gtrsim 0$, respectively. Here $p_{\rm F}=(3\pi^3 N)^{1/3}$ is the Fermi momentum of an assumed two-component free Fermi gas with $N$ atoms in the thermal equilibrium state. For later convenience, we also define the corresponding Fermi energy $\ep_{\rm F}=p_{\rm F}^2/(2m)$, as well as the Fermi temperature $T_{\rm F}=\ep_{\rm F}$.
\par
The left ($\alpha={\rm L}$) and right ($\alpha={\rm R}$) baths in Fig. \ref{fig1} are described by 
\begin{equation}
H_{\rm env}=\sum_{\alpha={\rm L},{\rm R}}\sum_{\bm{p},\sigma}
\xi^\alpha_{\bm p} c^{\alpha \dagger}_{\bm{p},\sigma} c^\alpha_{\bm{p},\sigma},
\label{Eq_H_env}
\end{equation}
where $c^\alpha_{\bm{p},\sigma}$ is an annihilation operator of a Fermi atom in the $\alpha$-bath, with the kinetic energy $\xi_{\bm p}^\alpha=\ep_{\bm p}-\mu_\alpha$, measured from the Fermi chemical potential $\mu_\alpha$. In this paper, we assume that both the baths are huge and can be described by free Fermi gases in the thermal equilibrium state at the common (environment) temperature $T_{\text{env}}$. Under these assumptions, the momentum distribution in each bath obeys the ordinary Fermi distribution function,
\begin{equation}
f(\xi_{\bm p}^\alpha)={1 \over e^{\xi_{\bm p}^\alpha/T_{\rm env}}+1}.
\label{eq.fermi}
\end{equation}
\par
The main system becomes in the non-equilibrium steady state by the couplings with the two baths $\alpha={\rm L,R}$. This coupling effect is described by \cite{PhysRevB.96.125206}
\begin{equation}
H_{\rm mix} \!=\!\! \sum_{\alpha={\rm L},{\rm R}} \! \sum_{i=1}^{N_{\rm t}} \!\sum_{\bm{p},\bm{q},\sigma}
\Bigl[
e^{i\mu_\alpha t} \!\Lambda_\alpha 
c^{\alpha\dagger}_{\bm{q},\sigma}
a_{\bm{p},\sigma}
e^{-i\bm{q}\cdot\bm{R}^\alpha_i} \!e^{i\bm{p}\cdot \bm{r}^\alpha_i}
+
{\rm h.c.}
\Bigr],
\label{Eq_H_mix}
\end{equation}
where $\Lambda_{\alpha={\rm L},{\rm R}}$ is a tunneling matrix element between the system and the $\alpha$-bath. In this paper, we set $\Lambda_{\rm L}=\Lambda_{\rm R}\equiv \Lambda$, for simplicity. In Eq. (\ref{Eq_H_mix}), the tunneling is assumed to occur between randomly distributing spatial positions ${\bm R}_i^\alpha$ in the $\alpha$-bath and positions ${\bm r}^\alpha_i$ in the system ($i=1,\cdots,N_{\rm t}~{\gg 1})$. Although the translational invariance of the system is then broken, this symmetry property will later recover by taking the spatial average over these tunneling positions \cite{PhysRevB.96.125206}. 
\par
In Eq. (\ref{Eq_H_mix}), the factor $e^{i\mu_\alpha t}$ describes the situation that the energy band in the $\alpha$-bath is filled up to $\mu_\alpha$ (at $T_{\rm env}=0$) \cite{PhysRevB.54.7366}, when the energy is measured from the bottom ($\varepsilon_{{\bm p}=0}=0$) of the energy band in the main system. When we write $\mu_{\rm L}=\mu+\delta\mu$ and $\mu_{\rm R}=\mu-\delta\mu$, the system becomes in the non-equilibrium state in the case of $\delta\mu\ne 0$. We call $\delta\mu$ the non-equilibrium parameter in this paper. Without loss of generality, we take $\delta\mu\ge 0$. In the non-equilibrium steady state, tunneling current flows from the left (pumping) bath to the system ($\equiv J_{\rm L}$) across the junction, as well as from the system to the right (decay) bath ($\equiv J_{\rm R}~(=J_{\rm L})$). However, the current does not flow in the main system, after taking the spatial average over the tunneling positions. 
\par
We briefly note that the system is in the thermal equilibrium state when $\delta\mu=0$, although the system-bath coupling still brings about non-zero quasi-particle damping.
\par
We also note that the system temperature is not well-defined in the non-equilibrium state. In the present case, the superfluid instability of the system is controlled by the temperature $T_{\rm env}$ of the thermal equilibrium baths. To emphasize this, we write the superfluid phase transition temperature as $T_{\rm env}^{\rm c}$ in this paper.
\par
\begin{figure*}[tbp]
\centering
\includegraphics[width=138mm]{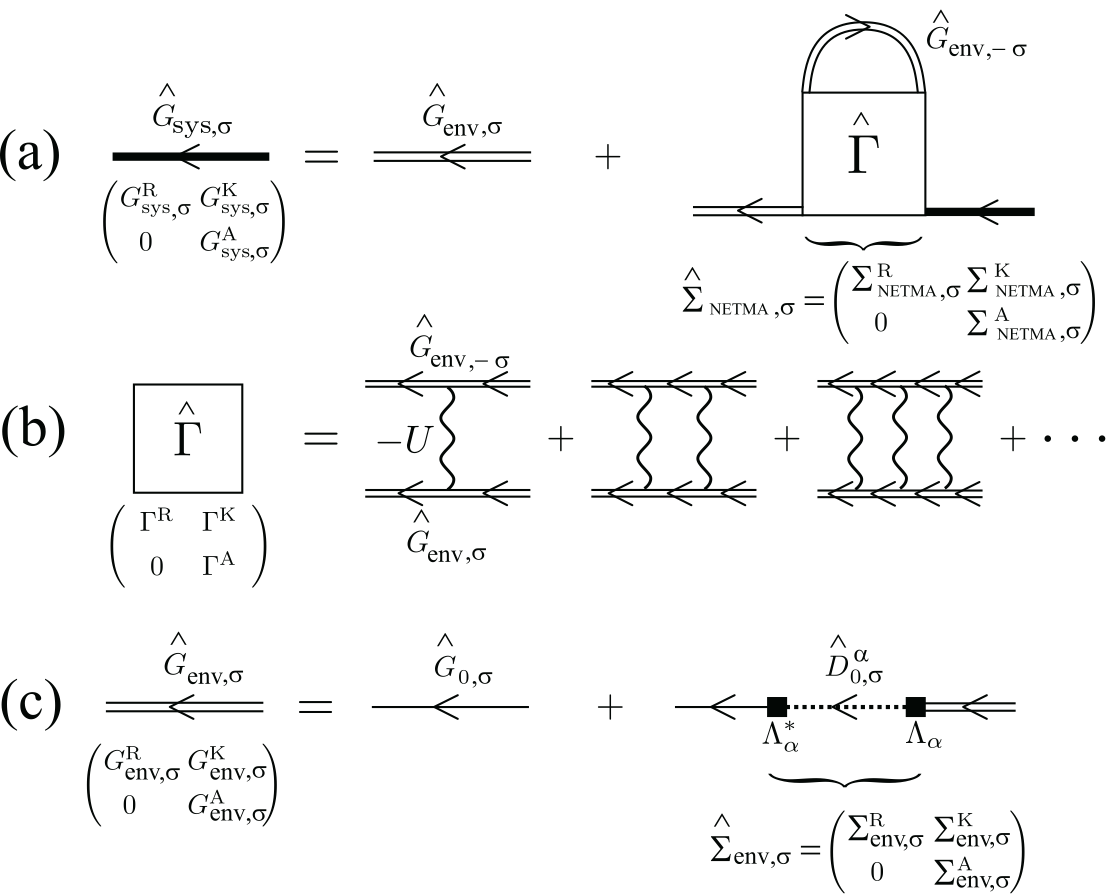}
\caption{(a) Dyson equation for $2\times 2$ matrix Keldysh Green's function ${\hat G}_{{\rm sys},\sigma}$ (thick solid line) in the main system. The self-energy ${\hat \Sigma}_{{\rm NETMA},\sigma}$ describes effects of pairing fluctuations in NETMA. (b) Particle-particle scattering matrix ${\hat \Gamma}$ in NETMA. The wavy line is the pairing interaction $-U$. The Keldysh Green's function ${\hat G}_{{\rm env},\sigma}$ (double solid line) involves effects of the two baths within the second-order Born approximation. (c) Dyson equation for ${\hat G}_{{\rm env},\sigma}$. The solid square denotes the tunneling $\Lambda_\alpha$ between the system and the $\alpha$-bath. ${\hat D}_{0,\sigma}^{\alpha={\rm L,R}}$ is the Keldysh Green's function in the $\alpha$-bath, given in Eq. (\ref{eq.DD}). ${\hat G}_{0,\sigma}$ is the single-particle propagator in the initial thermal equilibrium state in Eq. (\ref{Eq_G0}). In dealing with this Dyson equation, we take the spatial average over the tunneling positions ${\bf R}_i^\alpha$ and ${\bm r}_i^\alpha$, to recover the translational invariance.
}
\label{fig3} 
\end{figure*}
\par
\subsection{Keldysh Green's function in the main system}
\par
We extend TMA developed in the thermal equilibrium Fermi gas \cite{PhysRevB.66.024510,PhysRevA.80.033613} to the non-equilibrium steady state. For this purpose, we introduce the $2\times 2$ matrix Keldysh Green's function \cite{Rammer2007,Zagoskin,Stefanucci} in the main system,
\begin{eqnarray}
{\hat G}_{{\rm sys},\sigma}(\bm{p},\omega)
=
\left(
\begin{array}{cc}
G^{\rm R}_{{\rm sys},\sigma}(\bm{p},\omega) & 
G^{\rm K}_{{\rm sys},\sigma}(\bm{p},\omega) \\
0 & G^{\rm A}_{{\rm sys},\sigma}(\bm{p},\omega)
\end{array}
\right),
\label{eq.2}
\end{eqnarray}
where the superscripts `R', `A' and `K' represent the retarded, advanced, and Keldysh component, respectively. Strong coupling corrections to single-particle excitations can conveniently be described by the $2\times 2$ matrix self-energy,
\begin{equation}
{\hat \Sigma}_{{\rm NETMA},\sigma}(\bm{p}, \omega) \!=\!\!
\left(\!
\begin{array}{cc}
\Sigma_{{\rm NETMA},\sigma}^{\rm R}(\bm{p},\omega) & \Sigma_{{\rm NETMA},\sigma}^{\rm K}(\bm{p},\omega) \\
0 & \Sigma_{{\rm NETMA},\sigma}^{\rm A}(\bm{p},\omega)
\end{array}
\!\right) ,
\label{eq.2b}
\end{equation}
which appears in the Dyson equation \cite{Rammer2007,Zagoskin,Stefanucci} (see also Fig. \ref{fig3}(a)), 
\begin{align}
{\hat G}_{{\rm sys},\sigma}(\bm{p},\omega)&= {\hat G}_{{\rm env},\sigma}(\bm{p},\omega) \nonumber\\
&{\hskip -1cm} 
+{\hat G}_{{\rm env},\sigma}(\bm{p},\omega) {\hat \Sigma}_{{\rm NETMA},\sigma}(\bm{p},\omega) {\hat G}_{{\rm sys},\sigma}(\bm{p},\omega).
\label{Eq_Dyson_sys}
\end{align}
The Keldysh propagator ${\hat G}_{{\rm env},\sigma}$ has the same matrix structure as Eq. (\ref{eq.2}), but only involves effects of the system-bath coupling. Within the second-order Born approximation in terms of $\Lambda_{\alpha={\rm L,R}}$ (see Fig. \ref{fig3}(c)), it obeys the Dyson equation,
\begin{align}
{\hat G}_{{\rm env},\sigma}(\bm{p},\omega)&= {\hat G}_{0,\sigma}(\bm{p},\omega)  \nonumber\\
&+ 
{\hat G}_{0,\sigma}(\bm{p},\omega) 
{\hat \Sigma}_{{\rm env},\sigma}(\bm{p},\omega)
{\hat G}_{{\rm env},\sigma}(\bm{p},\omega),
\label{Eq_Dyson_env}
\end{align}
where we have taken the spatial average over the tunneling positions ${\bm R}_i^\alpha$ and ${\bm r}_i^\alpha$, and 
\begin{align}
{\hat G}_{0,\sigma}(\bm{p},\omega)&= \!
\left(
\begin{array}{cc}
G_{0,\sigma}^\text{R}(\bm{p},\omega) 
& G_{0,\sigma}^\text{K}(\bm{p},\omega)  \\
0& G_{0,\sigma}^\text{A}(\bm{p},\omega) 
\end{array}
\right)
\nonumber \\
&= \!
\left(
\begin{array}{cc}
{\displaystyle 1 \over \displaystyle \omega_+-\ep_{\bm{p}}} &
-2\pi i \delta(\omega-\ep_{\bm{p}}) \bigl[1-2f_{\rm ini}(\omega)\bigr] \\
0 &
{\displaystyle 1 \over \displaystyle\omega_--\ep_{\bm{p}}}
\end{array}
\right)
\label{Eq_G0}
\end{align}
is the bare Green's function in the initial non-interacting thermal equilibrium state at $t=-\infty$ \cite{Rammer2007,Zagoskin,Stefanucci}. Here, $\omega_\pm=\omega\pm i\delta$, where $\delta$ is an infinitesimally small positive number. $f_{\rm ini}(\omega)=1/[e^{\omega/T_{\rm ini}}+1]$ is the Fermi distribution function, where $T_{\rm ini}$ is the initial temperature. We will later find that the {\it final} non-equilibrium steady state actually looses the {\it initial} memory.
\par
The self-energy ${\hat \Sigma}_{{\rm env},\sigma}({\bm p},\omega)$ in Eq. (\ref{Eq_Dyson_env}) describes effects of the system-bath coupling, given by \cite{PhysRevB.96.125206},
\begin{align}
& {\hat \Sigma}_{{\rm env},\sigma}({\bm p},\omega) =
\left(
\begin{array}{cc}
\Sigma^{\rm R}_{{\rm env},\sigma}({\bm p},\omega)
 & \Sigma^{\rm K}_{{\rm env},\sigma}({\bm p},\omega) \\
0 & \Sigma^{\rm A}_{{\rm env},\sigma}({\bm p},\omega)
\end{array}
\right)
\nonumber \\
&=
\left(
\begin{array}{cc}
-2i\gamma  & -2i\gamma\left[\tanh\left(\frac{\omega-\mu_{\rm L}}{2T_{\rm env}}\right)+\tanh\left(\frac{\omega-\mu_{\rm R}}{2T_{\rm env}}\right)\right] \\[4pt]
0 & 2i\gamma
\end{array}
\right). 
\label{Eq_Sigma_env}
\end{align}
For the derivation of Eq. (\ref{Eq_Sigma_env}), see Appendix A. In Eq. (\ref{Eq_Sigma_env}), $\gamma=\pi N_{\rm t} \rho |\Lambda|^2$ is the quasi-particle damping, where $\rho~(=\rho_{\rm L}=\rho_{\rm R})$ is the single-particle density of state in the baths. (Note that we are setting $\Lambda_{\rm L}=\Lambda_{\rm R}=\Lambda$ in this paper.) For simplicity, we have employed the so-called wide-band limit approximation \cite{Stefanucci,RevModPhys.86.779}, where we have ignored the $\omega$-dependence, as well as the $\alpha(={\rm L,R})$-dependence, of the density of states $\rho$.
\par
The Dyson equation (\ref{Eq_Dyson_env}), as well as Eqs. (\ref{Eq_G0}) and (\ref{Eq_Sigma_env}), give
\begin{align}
&{\hat G}_{{\rm env},\sigma} (\bm{p},\omega)= \!
\left(
\begin{array}{cc}
G^{\rm R}_{{\rm env},\sigma}({\bm p},\omega)
 & G^{\rm K}_{{\rm env},\sigma}({\bm p},\omega) \\
0 & G^{\rm A}_{{\rm env},\sigma}({\bm p},\omega)
\end{array}
\right)
\nonumber
\\
=& \!
\left(
\begin{array}{cc}
\frac{\displaystyle 1}{\displaystyle \omega-\ep_{\bm{p}} +2i\gamma} &
\frac{\displaystyle -4i\gamma\left[1-f(\omega-\!\mu_{\rm L})-f(\omega-\!\mu_{\rm R})\right]}{\displaystyle [\omega-\ep_{\bm{p}}]^2 +4\gamma^2} \\[8pt]
0 &\frac{\displaystyle 1}{\displaystyle \omega-\ep_{\bm{p}} -2i\gamma}
\end{array}
\right). 
\label{Eq_G_env}
\end{align}
As mentioned previously, the Fermi distribution functions $f(\omega-\mu_{\alpha={\rm L,R}})$ in Eq. (\ref{Eq_G_env}) is nothing to do with the initial distribution $f_{\rm ini}(\omega)$ in Eq. (\ref{Eq_G0}). Instead, they are just the momentum distributions in the equilibrium pumping and decay baths, given in Eq. (\ref{eq.fermi}). This is because the initial condition is wiped out by the coupling with the pumping and decay baths \cite{Stefanucci, RevModPhys.86.779}. Because of the same reason, the main system in the non-equilibrium steady state is affected by, not the {\it initial} temperature $T_{\rm ini}$, but the (environmental) bath temperature $T_{\rm env}$, as seen in Eq. (\ref{Eq_G_env}). 
\par
We point out that two kinds of bath effects are involved in Eq. (\ref{Eq_G_env}): One is the quasi-particle damping $2\gamma$. The other is the deviation of the momentum distribution ($\equiv n_{{\bm p},\sigma}^{\rm env}$) from the Fermi distribution function \cite{Rammer2007,Zagoskin,Stefanucci}:
\begin{align}
n_{{\bm p},\sigma}^{\rm env}
&=
-i
\int_{-\infty}^\infty
{d\omega \over 2\pi}
G_{{\rm env},\sigma}^<({\bm p},\omega)
\nonumber
\\
&=
\int_{-\infty}^\infty
{d\omega \over 2\pi}
{2\gamma \over [\omega-\ep_{\bm p}]^2+4\gamma^2}
\left[f(\omega-\mu_{\rm L})+f(\omega-\mu_{\rm R})\right]
\nonumber
\\
&\simeq
{1 \over 2}
\left[f(\ep_{\bm p}-\mu_{\rm L})+f(\ep_{\bm p}-\mu_{\rm R})\right],
\label{eq.3}
\end{align}
where
\begin{align}
&G_{{\rm env},\sigma}^<({\bm p},\omega) \nonumber\\
=&
{1 \over 2}
\left[
G_{{\rm env},\sigma}^{\rm K}({\bm p},\omega)
-
G_{{\rm env},\sigma}^{\rm R}({\bm p},\omega)
+
G_{{\rm env},\sigma}^{\rm A}({\bm p},\omega)
\right]
\label{eq.lesser}
\end{align}
is the lesser Green's function \cite{Rammer2007,Zagoskin,Stefanucci}. In obtaining the last line in Eq. (\ref{eq.3}), we have taken the limit $\gamma\to 0$, for simplicity. In the thermal equilibrium case ($\mu_{\rm L}={\mu_{\rm R}}$), Eq. (\ref{eq.3}) is reduced to the expected ordinary Fermi distribution function $f(\varepsilon_{\bm p}-\mu)$. When the system is out of equilibrium ($\mu_{\rm L}>\mu_{\rm R}$), $n_{{\bm p},\sigma}^{\rm sys}$ in Eq (\ref{eq.3}) no longer equals the Fermi distribution function.
\par
\begin{widetext}
\par
\subsection{Non-equilibrium $T$-matrix approximation (NETMA)}
\par
We now evaluate the NETMA self-energy ${\hat \Sigma}_{{\rm NETMA},\sigma}({\bm p},\omega)$ in Eq. (\ref{eq.2b}). As in the equilibrium Matsubara formalism \cite{PhysRevB.66.024510,PhysRevA.80.033613}, pairing fluctuations in the Keldysh Green's function theory are also described by the ladder-type diagrams shown in Fig. \ref{fig3}(b). We explain how to sum up these diagrams in Appendix B, and immediately show the result here: 
\begin{align}
\Sigma^{\rm R}_{{\rm NETMA},\sigma}({\bm p},\omega) 
&=  \Bigl[\Sigma^{\rm A}_{{\rm NETMA},\sigma}({\bm p},\omega)\Bigr]^*
\nonumber
\\
&=
-\frac{i}{2}\sum_{\bm{q}} \int_{-\infty}^\infty \frac{d\nu}{2\pi}
\Bigl[\Gamma^\text{R}({\bm q},\nu) G_{{\rm env},-\sigma}^\text{K}({\bm q}-{\bm p},\nu-\omega) 
+
\Gamma^\text{K}({\bm q},\nu) G_{{\rm env},-\sigma}^\text{A}({\bm q}-{\bm p},\nu-\omega) \Bigr],
\label{Eq_Sigma_sys_R} \\
\Sigma^{\rm K}_{{\rm NETMA},\sigma}({\bm p},\omega) 
&=
-\frac{i}{2} \sum_{\bm{q}} \int_{-\infty}^\infty \frac{d\nu}{2\pi}
\Bigl[ 
\Gamma^{\rm A}({\bm q},\nu) G_{{\rm env},-\sigma}^\text{R}({\bm q}-{\bm p},\nu-\omega) 
\nonumber
\\
&\hskip 26mm+
\Gamma^{\rm R}({\bm q},\nu) G_{{\rm env},-\sigma}^\text{A}({\bm q}-{\bm p},\nu-\omega)
+
\Gamma^{\rm K}({\bm q},\nu) G_{{\rm env},-\sigma}^{\rm K}({\bm q}-{\bm p},\nu-\omega)
\Bigr].
\label{Eq_Sigma_sys_K}
\end{align}
Here, $\Gamma^{\rm R,A,K}({\bm q},\nu)$ are the retarded (R), advanced (A), and Keldysh (K) components of the particle-particle scattering matrices, that physically describe pairing fluctuations in the system. In NETMA, these are given by
\begin{eqnarray}
\left(
\begin{array}{cc}
\Gamma^{\rm R}({\bm q},\nu)&
\Gamma^{\rm K}({\bm q},\nu)\\
0&
\Gamma^{\rm A}({\bm q},\nu)
\end{array}
\right)
=
\left(
\begin{array}{cc}
\frac{\displaystyle -U}{\displaystyle 1+U\Pi^{\rm R}({\bm q},\nu)}
&
\frac{\displaystyle U^2\Pi^{\rm K}({\bm q},\nu)}{\displaystyle \bigl[1+U\Pi^{\rm R}({\bm q},\nu)\bigr]\bigl[1+U\Pi^{\rm A}({\bm q},\nu)\bigr]}\\[10pt]
0
&
\frac{\displaystyle -U}{\displaystyle 1+U\Pi^{\rm A}({\bm q},\nu)},
\end{array}
\right),
\label{Eq_Gamma}
\end{eqnarray}
where 
\begin{align}
\Pi^{\rm R}({\bm q},\nu) &= \bigl[\Pi^{\rm A}({\bm q},\nu)\bigr]^*
\nonumber
\\
&= \frac{i}{2} \sum_{\bm{p}} \int_{-\infty}^\infty \frac{d\omega}{2\pi}
\Bigl[
G^\text{R}_{{\text{env},\sigma}}({\bm p}+{\bm q}/2,\omega+\nu) G^\text{K}_{{\rm env},-\sigma}(-{\bm p}+{\bm q}/2,-\omega)
\nonumber
\\
&\hskip40mm+
G^\text{K}_{\text{env},\sigma}({\bm p}+{\bm q}/2,\omega+\nu) G^\text{R}_{{\rm env},-\sigma}(-{\bm p}+{\bm q}/2,-\omega)
\Bigr], 
\label{Eq_Pi_R}
\end{align}
\begin{align}
\Pi^\text{K}({\bm q},\nu)
&= 
\frac{i}{2} \sum_{\bm{p}} \int_{-\infty}^\infty \frac{d\omega}{2\pi}
\Bigl[
G^{\rm R}_{{\rm env},\sigma}({\bm p}+{\bm q}/2,\omega+\nu) G^\text{R}_{{\rm env},-\sigma}(-{\bm p}+{\bm q}/2,-\omega)
\nonumber
\\
&\hskip40mm+
G^{\rm A}_{{\rm env},\sigma}({\bm p}+{\bm q}/2,\omega+\nu) G^\text{A}_{{\rm env},-\sigma}(-{\bm p}+{\bm q}/2,-\omega)
\nonumber
\\
&\hskip40mm+
G^{\rm K}_{{\rm env},\sigma}({\bm p}+{\bm q}/2,\omega+\nu) G^\text{K}_{{\rm env},-\sigma}(-{\bm p}+{\bm q}/2,-\omega)
\Bigr], 
\label{Eq_Pi_K}
\end{align}
are the pair correlation functions.
\par
Substituting Eqs. (\ref{Eq_G_env}), (\ref{Eq_Sigma_sys_R}), and (\ref{Eq_Sigma_sys_K}), into the Dyson equation (\ref{Eq_Dyson_sys}), we obtain the NETMA Green's function ${\hat G}_{{\rm sys},\sigma}({\bm p},\omega)$ in the Keldysh space as 
\begin{align}
{\hat G}_{{\rm sys},\sigma}({\bm p},\omega)=
\left(
\begin{array}{cc}
\frac{\displaystyle 1}{\displaystyle \omega-\ep_{\bm{p}}+2i\gamma -\Sigma^{\rm R}_{{\rm NETMA},\sigma}(\bm{p},\omega)}
&
\frac{\displaystyle \Sigma^{\rm K}_{{\rm NETMA},\sigma}(\bm{p},\omega)+\Sigma^{\rm K}_{{\rm env},\sigma}(\bm{p},\omega)}{\displaystyle \bigl|\omega-\ep_{\bm{p}}+2i\gamma-\Sigma_{{\rm NETMA},\sigma}^\text{R}(\bm{p},\omega)\bigr|^2}
\\
0
&
\frac{\displaystyle 1}{\displaystyle \omega-\ep_{\bm{p}}-2i\gamma -\Sigma^{\rm A}_{{\rm NETMA},\sigma}(\bm{p},\omega)}
\end{array}
\right).
\label{Eq_G_sys}
\end{align}
\par
We note that, in the thermal equilibrium state, we actually only need to calculate the retarded components in Eqs. (\ref{Eq_Sigma_sys_R}), (\ref{Eq_Gamma}), (\ref{Eq_Pi_R}), and (\ref{Eq_G_sys}): The advanced components equal the complex conjugate of the retarded ones, e.g., $G^{\rm A}_{{\rm sys},\sigma}({\bm p},\omega)=[G^{\rm R}_{{\rm sys},\sigma}({\bm p},\omega)]^*$. The Keldysh components are also related to the retarded ones through the fluctuation-dissipation relations \cite{Rammer2007,Zagoskin,Stefanucci}. The latter example can been seen in the case of Eq. (\ref{Eq_G0}), as
\begin{equation}
G_0^{\rm K}({\bm p},\omega)=2i
{\rm Im}[G_0^{\rm R}({\bm p},\omega)]
\tanh \left( {\omega \over 2T_{\rm ini}} \right).
\label{eq.FDT}
\end{equation}
We will meet more examples in Sec. II.E. Such simple relations, however, no longer hold out of equilibrium. In this case, we need to treat the retarded and Keldysh components independently. The relation between the retarded and advanced components holds even out of equilibrium.
\par
\subsection{Equation for $T^{\rm c}_{\rm env}$ and single-particle quantities in NETMA}
\par
We determine the superfluid phase transition temperature $T^{\rm c}_{\rm env}$ by extending the theory developed by Kadanoff and Martin \cite{Kadanoff1961, Kadanoff1962} to the present non-equilibrium system. According to their theory, the system experiences the superfluid instability, when the retarded particle-particle scattering matrix $\Gamma^{\rm R}({\bm q},\nu)$ in Eq. (\ref{Eq_Gamma}) has a pole at ${\bm q}=0$ and a certain {\it real} value of the energy $\nu~(\equiv 2{\tilde \mu})$. This condition is given by
\begin{equation}
1+U\Pi^{\rm R}(\bm{q}=0,\nu=2{\tilde \mu})=0.
\label{Eq_TC}
\end{equation}
Because $\Pi^{\rm R}({\bm q},\nu)$ in Eq. (\ref{Eq_Pi_R}) is a complex function, Eq. (\ref{Eq_TC}) involves two equations, ${\rm Re}[\Gamma^{\rm R}(\bm{0},\nu=2{\tilde \mu})]^{-1}=0$ and ${\rm Im}[\Gamma^{\rm R}(\bm{0},\nu=2{\tilde \mu})]^{-1}=0$. Substituting Eq. (\ref{Eq_Gamma}) into the latter equation, we obtain
\begin{equation}
0=
\sum_{\bm{p},s=\pm 1} \int_{-\infty}^\infty \frac{d\omega}{2\pi} s
\frac
{
\tanh\left(\frac{\omega+s[\mu_{\rm L}-\tilde{\mu}]}{2T_{\rm env}^{\rm c}}\right) + 
\tanh\left(\frac{\omega+s[\mu_{\rm R}-\tilde{\mu}]}{2T_{\rm env}^{\rm c}} \right)
}
{\bigl[(\omega+\ep_{\bm{p}}-{\tilde \mu})^2 + 4\gamma^2 \bigr]
\bigl[(\omega-\ep_{\bm{p}}+{\tilde \mu})^2 + 4\gamma^2 \bigr] }.
\label{Eq_TC_Im}
\end{equation}
Because $\mu_{\rm L}=\mu+\delta\mu$ and $\mu_{\rm R}=\mu-\delta\mu$, Eq. \eqref{Eq_TC_Im} is satisfied only when ${\tilde \mu}=\mu$. Substituting this into the real part of Eq. \eqref{Eq_TC}, one obtains the $T_{\rm env}^{\rm c}$-equation as,
\begin{equation}
1=U\sum_{\bm{p}} \int_{-\infty}^\infty \frac{d\omega}{2\pi} \hspace{0.1cm}
\frac{2\gamma\omega \left[ \tanh\left(\frac{\omega+\delta\mu}{2T_\text{env}^\text{c}}\right)+
\tanh\left(\frac{\omega-\delta\mu}{2T_\text{env}^\text{c}}\right) \right]}
{\bigl[(\omega-\ep_{\bm{p}}+\mu)^2 +4\gamma^2\bigr]
 \bigl[(\omega+\ep_{\bm{p}}-\mu)^2 +4\gamma^2\bigr] }.
\label{Eq_TC_Re}
\end{equation}
\par
\begin{figure}[tb]
\centering
\includegraphics[width=85mm]{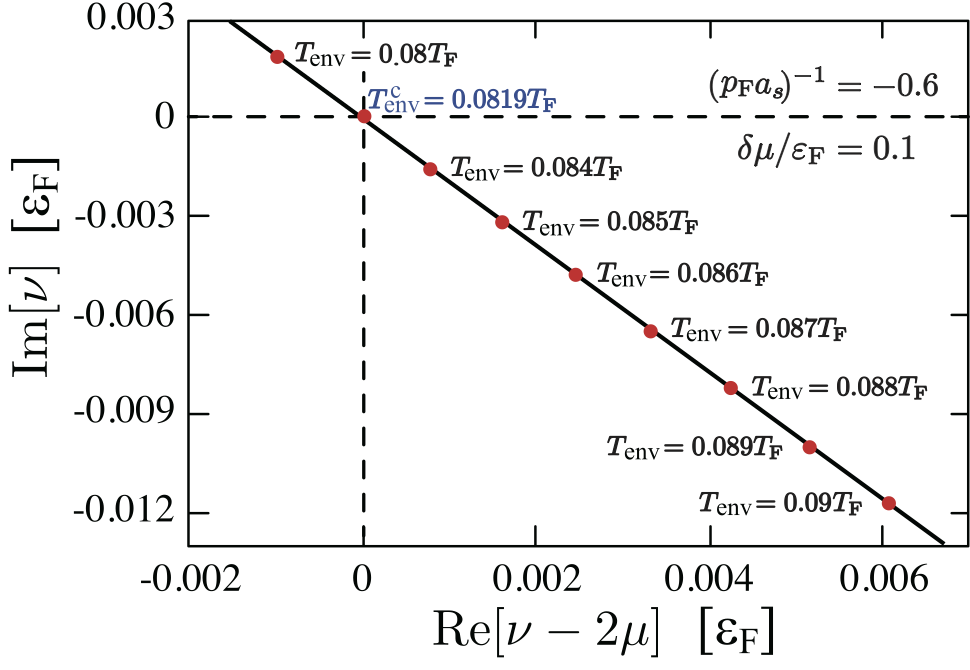}
\caption{Complex pole $\nu$ of the retarded particle-particle scattering matrix $\Gamma^{\rm R}({\bm q}=0,\nu)$ near $T_{\rm env}^{\rm c}$. We set $(p_{\rm F}a_s)^{-1}=-0.6$ and $\delta\mu/\varepsilon_{\rm F}=0.1$.}
\label{fig4} 
\end{figure}
\par
Above $T_{\rm env}^{\rm c}$ (normal state), $\Gamma^{\rm R}(\bm{q}=\bm{0},\nu)$ has a complex pole in the {\it lower}-half complex plain (${\rm Im}[\nu]<0$), as shown in Fig. \ref{fig4}. This means that when this fluctuation mode is excited, it finally damps out, that is, the system is stable. Figure \ref{fig4} also shows that the complex pole moves to the {\it upper} half plane (${\rm Im}[\nu]>0$) when $T_{\rm env}<T_{\rm env}^{\rm c}$, indicating the growth of this mode. This means the breakdown of the theory being based on the assumption of {\it stable} normal state \cite{Kadanoff1961,Kadanoff1962,AGD}. This indicates the superfluid instability at $T_{\rm env}^{\rm c}$. 
\par
As in the equilibrium BCS-BEC crossover theory \cite{PhysRevA.80.033613}, we  solve the $T^{\rm c}_{\rm env}$-equation (\ref{Eq_TC_Re}), together with the equation for the total number $N$ of Fermi atoms,
\begin{align}
N&=-i\sum_{\bm{p},\sigma}\int_{-\infty}^\infty \frac{d\omega}{2\pi} 
G^<_{{\rm sys},\sigma}(\bm{p},\omega)
\notag\\
&=
-2i\sum_{\bm{p}} \int_{-\infty}^\infty \frac{d\omega}{2\pi}
\frac{2i\gamma\bigl[f(\omega-\mu_{\rm L})+f(\omega-\mu_{\rm R})\bigr]
+
\Sigma_{{\rm NETMA},\sigma}^<({\bm p},\omega)}
{\bigl[\omega-\ep_{\bm{p}} -{\rm Re}[\Sigma^{\rm R}_{{\rm NETMA},\sigma}(\bm{p},\omega)]\bigr]^2 +
\bigl[2\gamma-{\rm Im}[\Sigma^{\rm R}_{{\rm NETMA},\sigma}(\bm{p},\omega)]\bigr]^2},
\label{Eq_number}
\end{align}
\end{widetext}
to determine $T_{\rm env}^{\rm c}$ and $\mu$ self-consistently. Here, 
\begin{align}
& G^<_{{\rm sys},\sigma}(\bm{p},\omega) \nonumber\\
=&{1 \over 2}\left[G^{\rm K}_{{\rm sys},\sigma}(\bm{p},\omega)-G^{\rm R}_{{\rm sys},\sigma}(\bm{p},\omega)+G^{\rm A}_{{\rm sys},\sigma}(\bm{p},\omega)\right]
\label{eq.G_less}
\end{align}
is the NETMA lesser Green's function, and 
\begin{align}
\Sigma_{{\rm NETMA},\sigma}^<(\bm{p},\omega)&=
{1 \over 2}\Bigl[ \Sigma_{{\rm NETMA},\sigma}^{\rm K}(\bm{p},\omega) \nonumber\\
&{\hskip -1.0cm}
-\Sigma_{{\rm NETMA},\sigma}^{\rm R}(\bm{p},\omega)+\Sigma_{{\rm NETMA},\sigma}^{\rm A}(\bm{p},\omega)
\Bigr].
\label{eq.sigma_less}
\end{align}
In the normal state above $T_{\rm env}^{\rm c}$, we only solve the number equation (\ref{Eq_number}), to determine $\mu$ for a given parameter set $(N, \delta\mu, (p_{\rm F}a_s)^{-1}, T_{\rm env})$. 
\par
Once $\mu$ is determined, the single-particle spectral weight (SW) $A({\bm p},\omega)$, as well as the single-particle density of state (DOS) $\rho(\omega)$, can be calculated from the retarded component of the NETMA Green's function as,
\begin{align}
&A(\bm{p},\omega) = -\frac{1}{\pi} 
{\rm Im}\bigl[G^{\rm R}_{{\rm sys},\sigma}(\bm{p},\omega)\bigr], 
\label{Eq_SW} 
\\
&\rho(\omega)= -\frac{1}{\pi} \sum_{\bm{p}} 
\text{Im}\bigl[G^{\rm R}_{{\rm sys},\sigma}(\bm{p},\omega)\bigr].  
\label{Eq_DOS}
\end{align} 
We also consider the photoemission spectrum (PES) $L({\bm p},\omega)$, which has recently become observable in cold Fermi gas physics \cite{Stewart2008,PhysRevLett.114.075301,Torma2016}. Because PES involves information about the occupation of SW, it is related to the lesser component as,
\begin{equation}
L(\bm{p},\omega) = -i p^2 G^<_{{\rm sys},\sigma}(\bm{p},\omega),
\label{Eq_PES}
\end{equation}
where we have suppressed an unimportant constant factor, for simplicity. In the thermal equilibrium state at $T_{\rm env}$, Eq. (\ref{Eq_PES}) reproduces the well-known expression \cite{Torma2016}, by using Eq. (\ref{eq.G_less}) and the fluctuation-dissipation relations among $G_{{\rm sys},\sigma}^{\rm R,K,A}$ \cite{Rammer2007,Zagoskin,Stefanucci},
\begin{equation}
L(\bm{p},\omega)=\pi p^2 A({\bm p},\omega)f(\omega).
\end{equation}
\par
\par
\subsection{NETMA in the thermal equilibrium limit}
\par
The coupled $T_{\rm env}^{\rm c}$-equation (\ref{Eq_TC_Re}) with the number equation (\ref{Eq_number}) in NETMA look very different from the corresponding equations in TMA \cite{PhysRevB.66.024510,PhysRevA.80.033613}. Before ending this section, we show that NETMA coincides with TMA in the thermal equilibrium limit $\delta\mu\to0$ and $\gamma\to+0$. 
\par
We first consider the condition for the superfluid instability in Eq. (\ref{Eq_TC}). When $\delta\mu=0$ and $\gamma\to+0$, the Green's function ${\hat G}_{{\rm env},\sigma}({\bm p},\omega)$ in Eq. (\ref{Eq_G_env}) is reduced to
\begin{align}
{\hat G}_{\!{\rm env},\sigma} (\bm{p},\omega)
\!=\!\!
\left(
\begin{array}{cc}
\! \frac{\displaystyle 1}{\displaystyle \omega_+-\ep_{\bm p}} &
\! -2i\delta(\omega\!-\!\ep_{\bm p})\left[1-2f(\omega\!-\!\mu)\right]\\
\!0 & 
\! \frac{\displaystyle 1}{\displaystyle \omega_--\ep_{\bm p}}
\end{array}
\right) . 
\label{Eq_G_env_eq}
\end{align}
Substituting this into Eq. (\ref{Eq_Pi_R}), and carrying out the $\omega$-integration, we obtain
\begin{equation}
\Pi^{\rm R}({\bm q},\nu)=
\sum_{\bm p}
\frac{1-f(\ep_{\bm{p}+\bm{q}/2}-\mu)-f(\ep_{-\bm{p}+\bm{q}/2}-\mu)}{\ep_{\bm{p}+\bm{q}/2}+\ep_{-\bm{p}+\bm{q}/2}-\nu_+}.
\label{eq.Pi_R0}
\end{equation}
Equation (\ref{eq.Pi_R0}) with $\nu=2\mu$ is just the same as the pair-correlation function \cite{PhysRevA.80.033613}, 
\begin{equation}
\Pi^{\rm M}({\bm q},\nu_n)=
\sum_{\bm p}
\frac{1-f(\ep_{\bm{p}+\bm{q}/2}-\mu)-f(\ep_{-\bm{p}+\bm{q}/2}-\mu)}{\ep_{\bm{p}+\bm{q}/2}+\ep_{-\bm{p}+\bm{q}/2}-2\mu-i\nu_n},
\label{eq.Pi_Matsu}
\end{equation}
in the Matsubara formalism at $\nu_n=0$ (where $\nu_n$ is the boson Matsubara frequency \cite{AGD}), when $\mu$ is interpreted as the Fermi chemical potential. Equation (\ref{Eq_TC}) is thus equivalent to the Thouless criterion \cite{Thouless1960}, $1=U\Pi^{\rm M}({\bm q}=0,\nu_n=0)$, in the thermal equilibrium limit. TMA uses this criterion, to determine the superfluid phase transition temperature \cite{PhysRevB.66.024510,PhysRevA.80.033613}.
\par
We also obtain the same conclusion for the $T_{\rm env}^{\rm c}$-equation (\ref{Eq_TC_Re}): Taking $\delta\mu\to0$, and $\gamma\to +0$ in this equation, we obtain
\begin{widetext}
\begin{eqnarray}
1
&=&
{U \over 2}\sum_{s=\pm 1}s\sum_{\bm p}
{1 \over \varepsilon_{\bm p}-\mu}
\int_{-\infty}^\infty {d\omega \over 2\pi}
\tanh
\left(
{\omega \over 2T_{\rm env}^{\rm c}}
\right)
\lim_{\gamma\to+0}
{2\gamma \over [\omega-s(\varepsilon_{\bm p}-\mu)]^2+(2\gamma)^2}
\nonumber
\\
&=&
{U \over 2}
\sum_{\bm p}
{1 \over \varepsilon_{\bm p}-\mu}
\tanh
\left(
{\varepsilon_{\bm p}-\mu \over 2T_{\rm env}^{\rm c}}
\right).
\label{eq.BCS}
\end{eqnarray}
\end{widetext}
This is just the ordinary BCS gap equation at the superfluid phase transition temperature \cite{Schrieffer1983}. TMA also uses this equation \cite{PhysRevB.66.024510,PhysRevA.80.033613}.
\par
We next consider the number equation (\ref{Eq_number}). In the thermal equilibrium limit, the Keldysh component in Eq. (\ref{Eq_G_env_eq}) satisfies the fluctuation-dissipation relation \cite{Rammer2007,Zagoskin,Stefanucci},
\begin{align}
&G^{\rm K}_{{\rm env},\sigma}({\bm p},\omega)  \nonumber\\
=& 
\left[
G^{\rm R}_{{\rm env},\sigma}({\bm p},\omega)-G^{\rm A}_{{\rm env},\sigma}({\bm p},\omega)
\right]
\tanh
\left(
{\omega-\mu \over 2T_{\rm env}}
\right).
\label{FDT_G}
\end{align}
Using this, we find that $\Pi^{\rm K}({\bm q},\nu)$ in Eq. (\ref{Eq_Pi_K}), as well as $\Gamma^{\rm K}({\bm q},\nu)$ in Eq. (\ref{Eq_Gamma}), obey the following relations, respectively,
\begin{align}
\Pi^{\rm K}({\bm q},\nu)
&=
\left[
\Pi^{\rm R}({\bm q},\nu)-\Pi^{\rm A}({\bm q},\nu)
\right]
\coth
\left(
{\nu-2\mu \over 2T_{\rm env}}
\right),
\label{FDT_PI}
\\
\Gamma^{\rm K}({\bm q},\nu)
&=
\left[
\Gamma^{\rm R}({\bm q},\nu)-\Gamma^{\rm A}({\bm q},\nu)
\right]
\coth
\left(
{\nu-2\mu \over 2T_{\rm env}}
\right).
\label{FDT_GAMMA}
\end{align}
Substituting Eqs. (\ref{FDT_G}) and (\ref{FDT_GAMMA}) into Eq. (\ref{Eq_Sigma_sys_K}), one finds,
\begin{align}
& \Sigma^{\rm K}_{{\rm NETMA},\sigma}({\bm p},\omega) \nonumber\\
=&
\left[
\Sigma^{\rm R}_{{\rm NETMA},\sigma}({\bm p},\omega)
-\Sigma^{\rm A}_{{\rm NETMA},\sigma}({\bm p},\omega)
\right]
\tanh
\left(
{\omega-\mu \over 2T_{\rm env}}
\right).
\label{FDT_SIGMA}
\end{align}
Then, the lesser component $\Sigma^<_{{\rm NETMA},\sigma}({\bm p},\omega)$ of the self-energy in Eq. (\ref{eq.sigma_less}) in the thermal equilibrium limit can be expressed by using the retarded component as,
\begin{equation}
\Sigma^<_{{\rm NETMA},\sigma}({\bm p},\omega)=
-2i
{\rm Im}[\Sigma^{\rm R}_{{\rm NETMA},\sigma}({\bm p},\omega)]
f(\omega-\mu).
\label{FDT_LESS}
\end{equation}
Thus, the number equation (\ref{Eq_number}) in the thermal equilibrium limit is written as,
\begin{equation}
N=2\int_{-\infty}^\infty d\omega \rho(\omega+\mu)f(\omega),
\label{eq_num_eq}
\end{equation}
where DOS $\rho(\omega)$ is given in Eq. (\ref{Eq_DOS}). Using the fluctuation-dissipation relations in Eqs. (\ref{FDT_G}) and (\ref{FDT_GAMMA}), one finds that the retarded self-energy $\Sigma^{\rm R}_{{\rm NETMA},\sigma}({\bm p},\omega+\mu)$ in Eq. (\ref{Eq_Sigma_sys_R}) in the thermal equilibrium limit has the form,
\begin{align}
& \Sigma^{\rm R}_{{\rm NETMA},\sigma}(\bm{p},\omega+\mu) \nonumber\\
=& 
-\sum_{\bm p}
\int_{-\infty}^\infty
{d\nu \over \pi}
{\rm Im}[\Gamma^{\rm R}({\bm q},\nu+2\mu)]
{n_{\rm B}(\nu)-f(\ep_{\bm p}-\mu)
\over \omega_+-\nu+\ep_{{\bm q}-{\bm p}}-\mu}.
\label{eq.Sigma_eq}
\end{align}
Here, $n_{\rm B}(\nu)=[e^{\nu/T_{\rm env}}-1]^{-1}$ is the Bose distribution function, and $\Pi^{\rm R}({\bm q},\nu)$ involved in $\Gamma^{\rm R}({\bm q},\nu)$ (see Eq. (\ref{Eq_Gamma})) is given in Eq. (\ref{eq.Pi_R0}). Equation (\ref{eq.Sigma_eq}) coincides with the TMA self-energy \cite{PhysRevA.80.033613}. The number equation (\ref{eq_num_eq}) is thus the same as that in TMA.
\par
The above discussions show that NETMA is a natural extension of TMA to the non-equilibrium steady state.
\par
\begin{figure}[t]
\centering
\includegraphics[width=86mm]{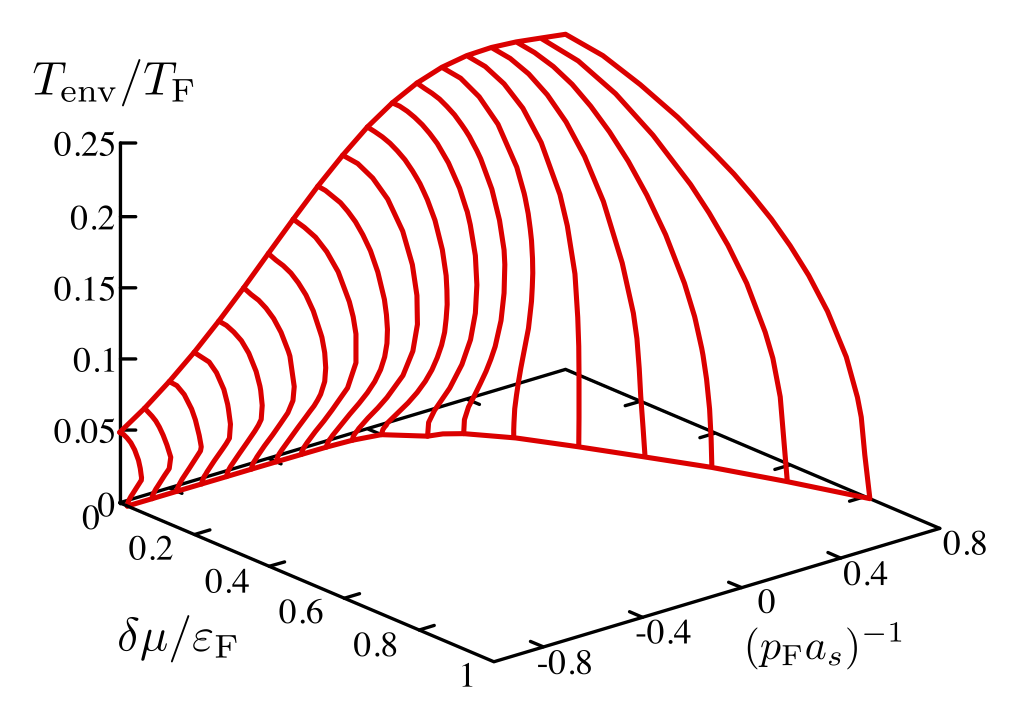}
\caption{Calculated superfluid phase transition temperature $T_{\rm env}^{\rm c}$ in a two-component non-equilibrium Fermi gas, as a function of the interaction strength $(p_{\rm F}a_s)^{-1}$ and the non-equilibrium parameter $\delta\mu$. We set $\gamma/\ep_{\rm F}=0.01$. This value is also used in the following figures. $\ep_{\rm F}=p_{\rm F}^2/(2m)$ and $T_{\rm F}$ are, respectively, the Fermi energy and the Fermi temperature in an assumed free Fermi gas with $N=p_{\rm F}^3/(3\pi^2)$ fermions in the equilibrium state.}
\label{fig5} 
\end{figure}
\par
\section{Strong-coupling properties of a Fermi gas in the non-equilibrium steady state}
\par
In this section, we consider the superfluid phase transition, as well as single-particle excitations, in the BCS-BEC crossover regime of a Fermi gas in the non-equilibrium steady state. As a typical many-body phenomenon discussed in the thermal equilibrium case, we pick up the pseudogap phenomenon.
\par
\subsection{Superfluid phase transition temperature $T_{\rm env}^{\rm c}$}
\par
Figure \ref{fig5} shows the superfluid phase transition temperature $T_{\rm env}^{\rm c}$ in the non-equilibrium steady state of an ultracold Fermi gas. When the non-equilibrium parameter $\delta\mu$ vanishes, the system is in the thermal equilibrium state. In this limiting case, although effects of the quasi-particle damping $\gamma$ involved in ${\hat \Sigma}_{{\rm env},\sigma}$ in Eq. (\ref{Eq_Sigma_env}), which comes from the system-bath coupling, still affects $T_{\rm env}^{\rm c}$, the interaction dependence of $T_{\rm env}^{\rm c}$ is essentially the same as the ordinary BCS-BEC crossover behavior known in the thermal equilibrium state \cite{Strinati,PhysRevA.80.033613}: Starting from the weak-coupling BCS regime ($(p_{\rm F}a_s)^{-1}\lesssim -1$), $T_{\rm env}^{\rm c}$ gradually increases with increasing the interaction strength, to approach a constant value ($T_{\rm c}\simeq 0.218T_{\rm F}$) \cite{NSR,Randeria1995} in the strong-coupling BEC regime ($(p_{\rm F}a_s)^{-1}\gg +1$). 
\par
When the system is in the non-equilibrium steady state ($\delta\mu>0$), Fig. \ref{fig5} shows that $T^{\rm c}_{\rm env}$ decreases from the equilibrium value ($\equiv T_{\rm env}^{{\rm c}0}$). When the non-equilibrium parameter $\delta\mu$ exceeds a certain value, the superfluid instability no longer occurs.
\par
This suppression of $T_{\rm env}^{\rm c}$ is because the system-bath coupling in the non-equilibrium state has a similar effect to the temperature in the thermal equilibrium state: When $\delta\mu=0$ at $T_{\rm env}=0$, the main system is in the thermal equilibrium state with the Fermi energy $\varepsilon_{\rm F}=\mu$. Starting from this, when $\delta\mu>0$, the pumping bath supplies atoms to the system in the energy region $\mu\lesssim \omega\lesssim \mu+\delta\mu$, and the decay bath absorbs atoms from the system in the region $\mu-\delta\mu\lesssim \omega\lesssim \mu$. As a result, the Fermi surface of the main system at $\omega=\mu$ is smeared out within the energy range $\mu-\delta\mu\lesssim \omega\lesssim \mu+\delta\mu$. This smearing effect is similar to the thermal effect on the Fermi surface in the thermal equilibrium state. Since the latter effect is known to be unfavorable to the superfluid instability, the increase of the non-equilibrium parameter $\delta\mu$ also suppresses $T_{\rm env}^{\rm c}$.
\par
Although the above discussion explains the suppression of $T_{\rm env}^{\rm c}$ seen in Fig. \ref{fig5}, this figure also shows that the detailed $\delta\mu$-dependence of $T_{\rm env}^{\rm c}$ in the BCS-unitary regime ($(p_Fa_s)^{-1}\lesssim 0$) is quite different from that in the BEC regime ($(p_Fa_s)^{-1}\gesim 0$). In the latter, $T^{\rm c}_{\rm env}$ {\it monotonically} decreases with increasing $\delta\mu$. In the former case, on the other hand, $T_{\rm env}^{\rm c}$ exhibits {\it re-entrant} behavior. To explain the reason for this difference, we separately consider the BCS-unitary regime and BEC regime in the followings.
\par
\begin{figure*}[tbp]
\centering
\includegraphics[width=17.8cm]{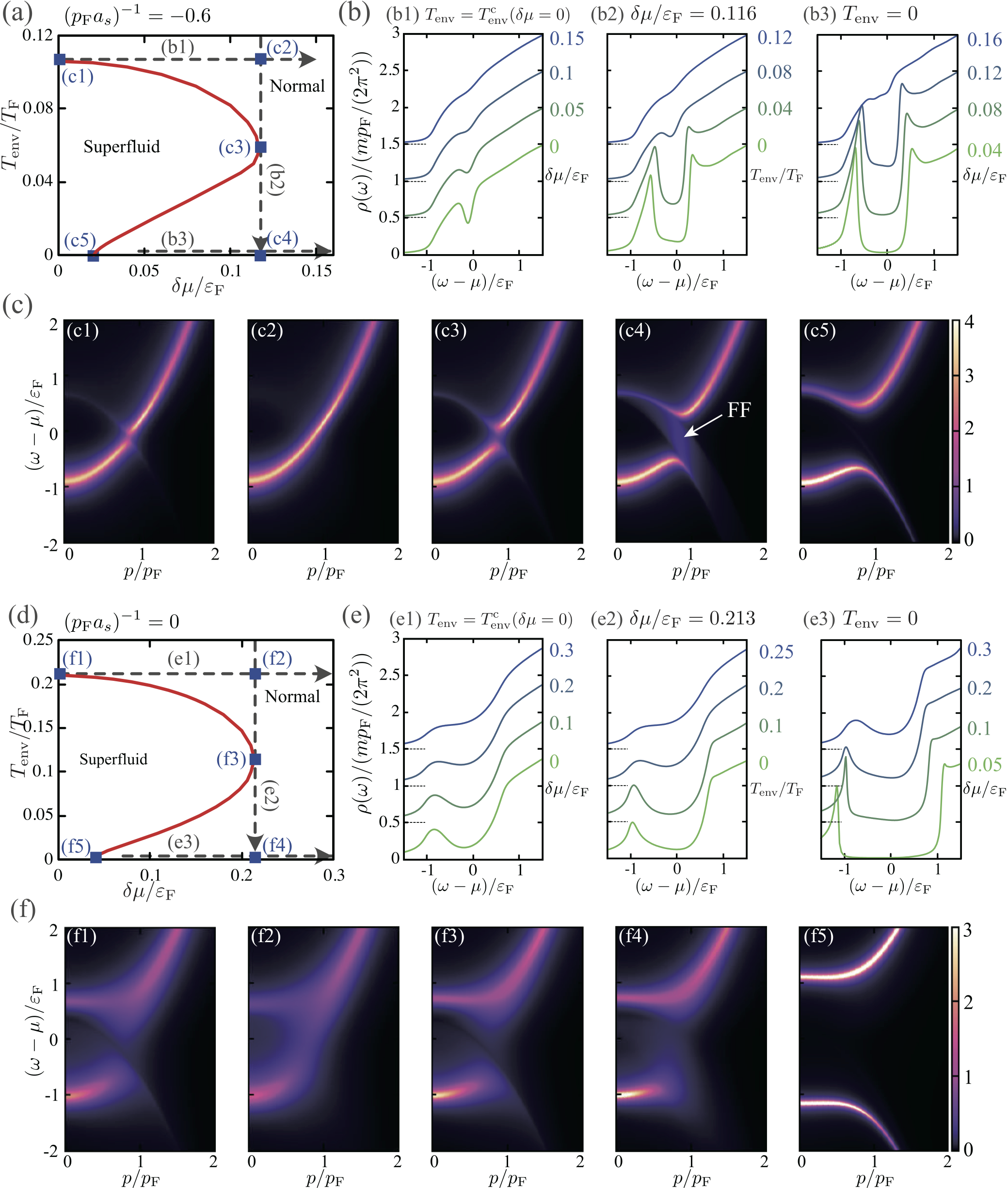}
\caption{Single-particle properties of a non-equilibrium Fermi gas in the BCS regime at $(p_{\rm F}a_s)^{-1}=-0.6$. (a) $T_{\rm env}^{\rm c}$. (b) Single-particle density of states (DOS) $\rho(\omega)$. Each panel shows the result along the path (b1)-(b3) in panel (a). We fix $T_{\rm env}$ in panels (b1) and (b3), and fix $\delta\mu$ in panel (b2). In these panels, we offset the results. The short horizontal line near each result is at $\rho(\omega)=0$. (c) Intensity of single-particle spectral weight (SW) $A(\bm{p},\omega)$, normalized by $\ep_{\rm F}^{-1}$. Each panel corresponds to the case at (c1)-(c5) in panel (a). The broad downward spectral structure `FF' in panel (c4) is associated with FFLO-type pairing fluctuations (see Sec. III.B.4). (d)-(f) Same plots as panels (a)-(c), for $(p_{\rm F}a_s)^{-1}=0$ (unitarity limit).
}
\label{fig6} 
\end{figure*}
\par
\subsection{Non-equilibrium effects in BCS-unitary regime}
\par
\par
\subsubsection{Single-particle excitations and pseudogap phenomenon}
\par
Figure \ref{fig6}(a) shows $T_{\rm env}^{\rm c}$ and non-equilibrium effects, when $(p_{\rm F}a_s)^{-1}=-0.6$ (BCS regime). A similar re-entrant behavior of $T_{\rm env}^{\rm c}$ is also obtained at the unitarity $(p_{\rm F}a_s)^{-1}=0$, as shown in Fig. \ref{fig6}(d). In order to examine the interplay between non-equilibrium and strong-coupling effects in these cases, we show DOS $\rho(\omega)$ and SW $A({\bm p},\omega)$ in panels (b) and (e), and panels (c) and (f), respectively.
\par
We find in Fig \ref{fig6}(b1) that the pseudogap structure in DOS (dip around $\omega-\mu=0$) \cite{PhysRevA.80.033613} gradually becomes obscure with increasing $\delta\mu$. We also find from Fig. \ref{fig6}(c1) and (c2) that the coupling of the particle branch ($\omega-\mu=\varepsilon_{\bm p}-\mu$) with the hole branch ($\omega-\mu=-[\varepsilon_{\bm p}-\mu]$), which is characteristic of the pseudogap phenomenon \cite{Levin2005,Strinati,PhysRevB.66.024510,PhysRevA.80.033613}, is weakened, when the system goes out of equilibrium. 
\par
Figure \ref{fig6}(b3) shows that the pseudogap also becomes obscure, as one moves along the path (b3) drawn in Fig. \ref{fig6}(a). At this temperature ($T_{\rm env}=0$), panels (c4) and (c5) indicate that the particle-hole coupling also becomes weak and the gap size shrinks with increasing $\delta\mu$ ((c5)$\to$(c4)). Furthermore, the same non-equilibrium effects on the pseudogap phenomenon also occur at the unitarity ($(p_{\rm F}a_s)^{-1}=0$): In panels (e1) and (e3), the dip structure gradually becomes less remarkable with increasing $\delta\mu$ along the paths (e1) and (e3) in Fig. \ref{fig6}(d), respectively. The weakening of the particle-hole coupling is also found from panels (f1)$\to$(f2) and (f5)$\to$(f4). 
\par
To understand these non-equilibrium effects, we recall that the essence of the pseudogap phenomenon in the thermal equilibrium state can be understood by using the so-called static approximation to pairing fluctuations \cite{PhysRevA.80.033613,Levin2005}. This approximation is also applicable to the non-equilibrium state: Assuming that pairing fluctuations in the present case are  enhanced around $\bm{q}=0$ and $\nu=2\mu$ near $T_{\rm env}^{\rm c}$ (Note that $\Gamma^\text{R}({\bm q}=0,\omega=2\mu)$ diverges at $T_\text{env}^\text{c}$.), we approximate the retarded component $\Sigma^{\rm R}_{{\rm NETMA},\sigma}(\bm{p},\omega)$ of the NETMA self-energy in Eq. (\ref{Eq_Sigma_sys_R}) to
\begin{equation}
\Sigma^{\rm R}_{{\rm NETMA},\sigma}(\bm{p},\omega)
\simeq
-\Delta_{\rm pg}^2 G_{{\rm env},-\sigma}^{\rm A}(-\bm{p},-\omega+2\mu).
\label{Eq_static}
\end{equation}
Here, 
\begin{equation}
\Delta_{\rm pg}^2 = \frac{i}{2}\sum_{\bm{q}} \int_{-\infty}^\infty \frac{d\nu}{2\pi} \Gamma^{\rm K}(\bm{q},\nu) 
\end{equation}
is the pseudogap parameter $\Delta_{\rm pg}$ \cite{PhysRevA.80.033613,Levin2005}, describing effects of pairing fluctuations. Substituting Eq. (\ref{Eq_static}) into the retarded component of Eq. (\ref{Eq_G_sys}), we obtain
\begin{align}
G^\text{R}_\text{sys}(\bm{p},\omega) 
&\simeq
 \frac{\omega-\mu+2i\gamma +\xi_{\bm{p}}}{\bigl[\omega-\mu+2i\gamma\bigr]^2-\bigl[\xi_{\bm{p}}^2+\Delta_{\rm pg}^2\bigr]}
\nonumber
\\
&=
{1 \over \displaystyle
(\omega-\mu)+2i\gamma-\xi_{\bm p}
-{\Delta_{\rm pg}^2 \over (\omega-\mu)+2i\gamma+\xi_{\bm p}
}},
\label{Eq_app_GR}
\end{align}
where $\xi_{\bm{p}}=\ep_{\bm{p}}-\mu$. When we ignore the damping parameter $\gamma$ for simplicity, the first line in Eq. (\ref{Eq_app_GR}) has the same form as the diagonal component of the mean-field BCS Green's function \cite{Schrieffer1983}, where the order parameter $\Delta$ is now replaced by the pseudogap parameter $\Delta_\text{pg}$. Thus, Eq. (\ref{Eq_app_GR}) gives a gapped DOS with the energy gap $2|\Delta_{\rm pg}|$. The second line in Eq. (\ref{Eq_app_GR}) also explains the coupling between the particle branch ($\omega-\mu=\xi_{\bm p}$) and the hole branch ($\omega-\mu=-\xi_{\bm p}$) with the coupling strength $\Delta_{\rm pg}^2$.
\par
In this scheme, the pseudogap phenomenon is simply characterized by the pseudogap parameter $\Delta_{\rm pg}$, describing the strength of pairing fluctuations. Because they becomes strong near the superfluid instability, it is usually believed that the pseudogap phenomenon is remarkable near $T_{\rm env}^{\rm c}$.
\par
However, Fig. \ref{fig6}(b2) shows that the pseudogap in DOS continues to develop with decreasing $T_{\rm env}$, even after passing through the superfluid phase transition point at (c3) shown in Fig. \ref{fig6}(a). Figures \ref{fig6}(c2)-(c4) also show the monotonic development of the level repulsion associated with the particle-hole coupling along this path. Furthermore, the same tendency is also seen in the unitary limit, as shown in Fig. \ref{fig6}(e2) (DOS), as well as in Figs. \ref{fig6}(f2)-(f4) (SW). 
\par
We note that the pseudogap phenomenon in the BCS-BEC crossover region is usually attributed to pairing fluctuations associated with the superfluid phase transition. Then, this many-body phenomenon should be the most remarkable at $T_{\rm env}^{\rm c}$. Thus, the above results imply that the remarkable pseudogap seen around (c4) and (f4) in Figs. \ref{fig6}(a) and (d) originate from different kinds of pairing fluctuations that are nothing to do with the superfluid state realized below $T_{\rm env}^{\rm c}$. We will explain detailed of this point in Sec. III.B.2.
\par
\begin{figure}[t]
\centering
\includegraphics[width=8.6cm]{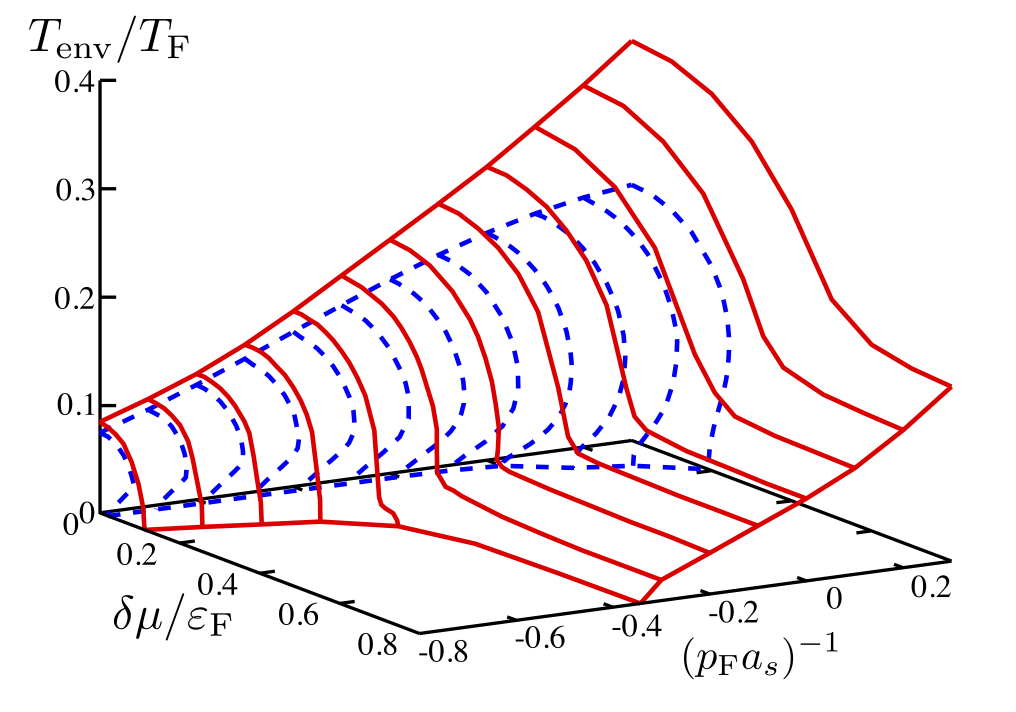}
\caption{Pseudogap temperature $T^*_{\rm env}$ (solid line) which is determined as the temperature below which a dip appears in DOS. The dotted line is $T^{\rm c}_{\rm env}$ shown in Fig. \ref{fig5}.}
\label{fig7} 
\end{figure}
\par
\par
\begin{figure}[t]
\centering
\includegraphics[width=7.2cm]{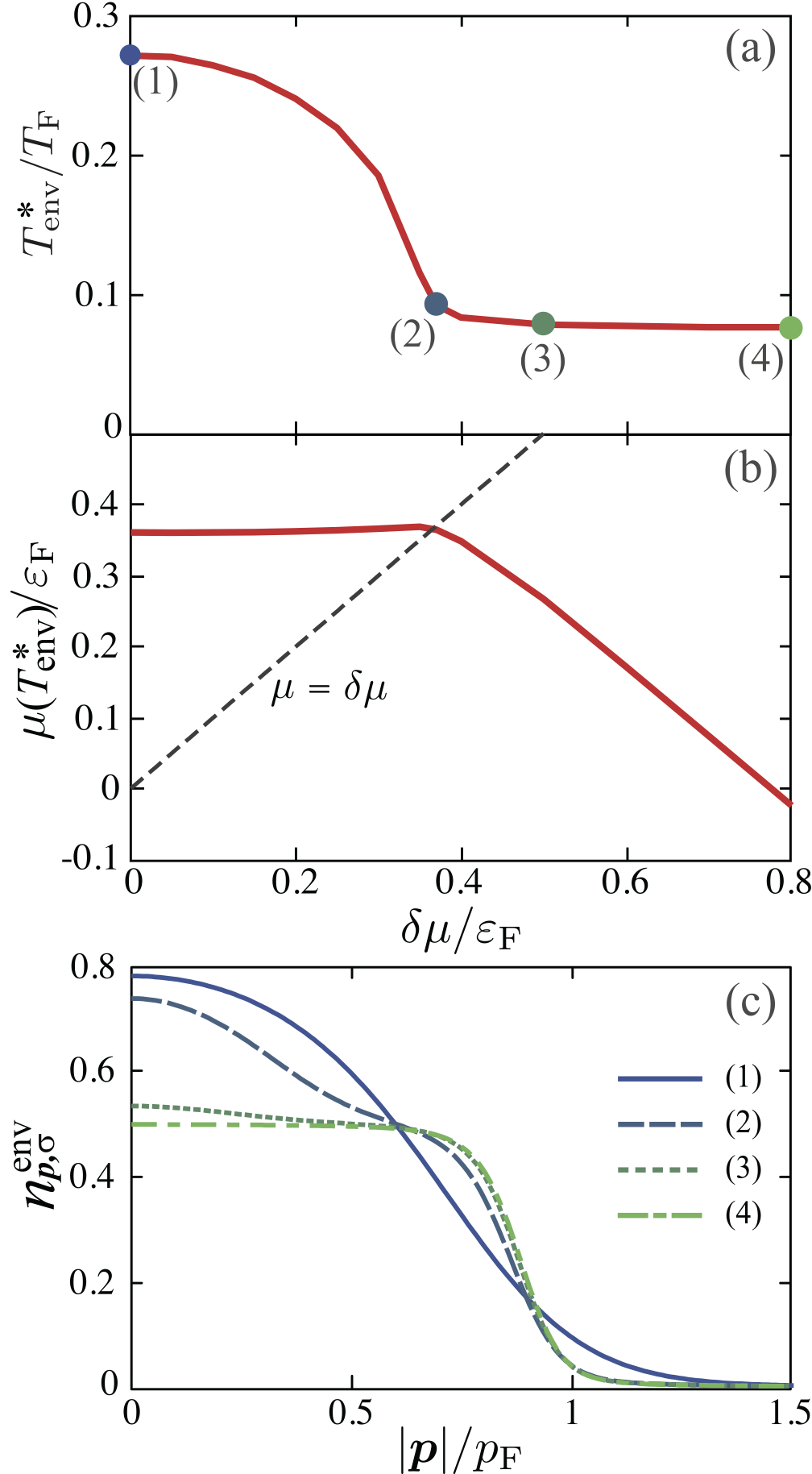}
\caption{
(a) Pseudogap temperature $T_{\rm env}^*$ at the unitarity ($(p_{\rm F}a_s)^{-1}=0$). (b) $\mu(T_{\rm env}^*)$ as a function of the non-equilibrium parameter $\delta\mu$. The dashed line is $\mu=\delta\mu$. (c) Atomic momentum distribution $n_{{\bm p},\sigma}^{\rm env}(T_{\rm env}^*)$ at (1)-(4) in panel (a).
}
\label{fig8} 
\end{figure}
\par
To specify the region where the pseudogapped DOS exists in the non-equilibrium steady state, we introduce the pseudogap temperature $T^*_{\rm env}$ \cite{PhysRevA.80.033613}, which is defined as the temperature below which a dip appears in DOS. Although this definition somehow involves ambiguity because the pseudogap is a crossover phenomenon without being accompanied by any phase transition, we still expect that $T^*_{\rm env}$ gives useful information about pairing fluctuations out of equilibrium. 
\par
Figure \ref{fig7} shows $T^*_{\rm env}$ in the BCS-unitary regime. The region enclosed by $T_{\rm env}^*$ and $T^{\rm c}_{\rm end}$ is the pseudogap regime, where the pseudogapped DOS is obtained. In the BCS regime when $(p_{\rm F}a_s)^{-1}\lesssim -0.4$, we find that the pseudogap temperature $T_{\rm env}^*$ does not exhibit re-entrant behavior, but {\it monotonically} decreases with increasing $\delta\mu$. This again implies that dominant pairing fluctuations around $T_{\rm env}=0$ are different from pairing fluctuations that are directly related to the superfluid phase transition.  
\par
In the unitary regime ($(p_{\rm F}a_s)^{-1}\gesim -0.4$), Fig. \ref{fig7} shows that $T_{\rm env}^*$ is almost $\delta\mu$-independent when $\delta\mu/\varepsilon_{\rm F}\gesim 0.4$. This is because the decay bath extracts atoms from the system in the energy region $\mu-\delta\mu\lesssim \omega\lesssim \mu$, so that atoms in all the energy region $0\lesssim\omega\lesssim \mu$ in the system can leak into the decay bath, when $\delta\mu\gesim\mu$. As a result, the $\delta\mu$-dependence of the atomic momentum distribution $n_{{\bm p},\sigma}^{\rm env}$, given in the first line in Eq. (\ref{eq.3}), becomes weak when $\delta\mu\gesim\mu$. Because the modification of the momentum distribution $n_{{\bm p},\sigma}^{\rm env}$ by the system-bath coupling is crucial for system properties, the $\delta\mu$-dependence of $T_{\rm env}^*$ becomes weak for $\delta\mu\gesim\mu$. 
\par
To confirm this explanation,  we compare $T_{\rm env}^*$ with $\mu(T_{\rm env}^*)$ at the unitarity $(p_{\rm F}a_s)^{-1}=0$, in Figs. \ref{fig8}(a) and (b). We find that the $\delta\mu$-dependence of $T_{\rm env}^*$ becomes weak when $\delta\mu$ becomes larger than $\mu(T_{\rm env}^*)$, as expected. Figure \ref{fig8}(c) also shows that the momentum distribution $n^{\rm env}_{{\bm p},\sigma}$ is almost the same in the cases of (3) and (4).
\par
\begin{figure}[t]
\centering
\includegraphics[width=8.6cm]{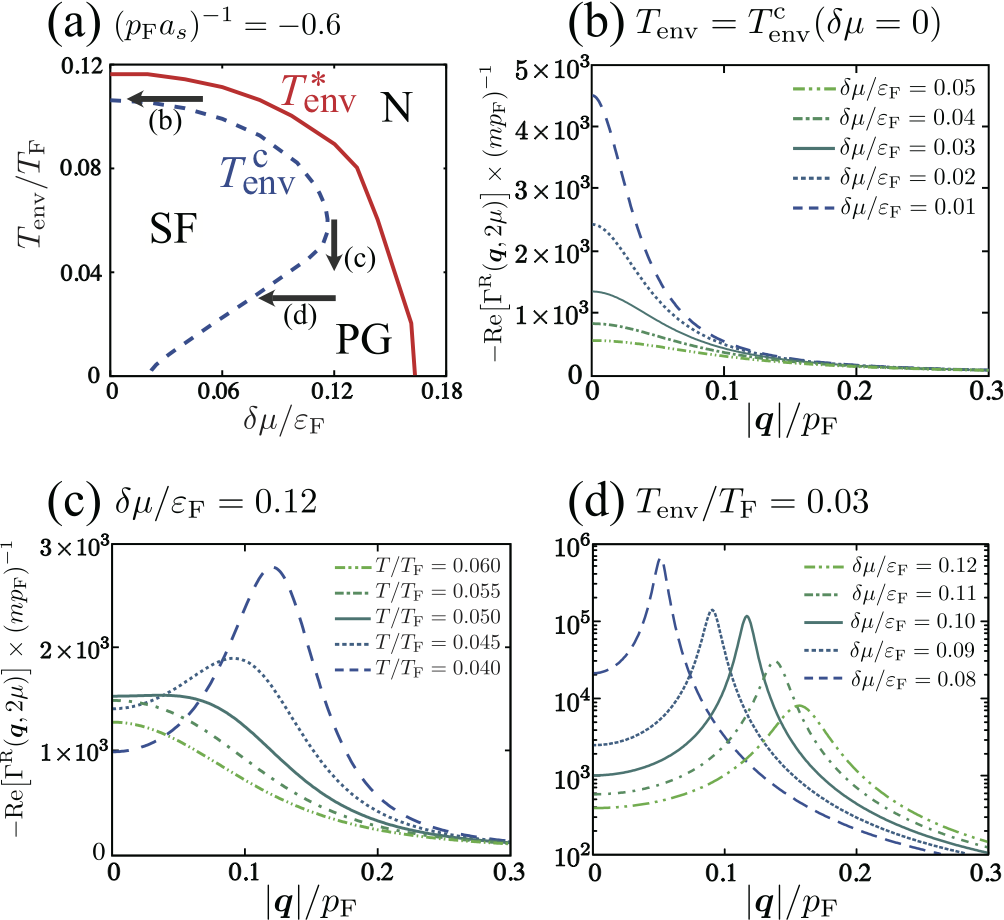}
\caption{(a) Phase diagram in the BCS regime when $(p_{\rm F}a_s)^{-1}=-0.6$. SF: Superfluid state. N: Normal state. PG: Pseudogap regime. (b)-(d) $-{\rm Re}\bigl[\Gamma^{\rm R}(\bm{q},\nu=2\mu)\bigr]$ in NETMA, as a function of the momentum $|\bm{q}|$. Each panel shows the result along the path (b)-(d) shown in panel (a).}
\label{fig9} 
\end{figure}
\par
\subsubsection{Enhancement of FFLO-type pairing fluctuations in the non-equilibrium BCS-unitary regime}
\par
We discuss the two phenomena obtained in Sec. III.B.1: (1) The pseudogap continues to develop as one moves along the paths (b2) and (e2) in Figs. \ref{fig6}(a) and (d), even after passing through the phase transition points (c3) and (f3). (2) The pseudogap temperature $T_{\rm env}^*$ does not exhibit re-entrant behavior in the BCS regime where $T_{\rm env}^{\rm c}$ does.
\par
Figures \ref{fig9}(b)-(d) show the real part $-{\rm Re}\bigl[\Gamma^{\rm R}(\bm{q},\nu=2\mu)\bigr]$ of the retarded particle-particle scattering matrix in Eq. (\ref{Eq_Gamma}). We recall that NETMA determines the superfluid phase transition temperature from the condition that it diverges at ${\bm q}=0$ (see Eq. (\ref{Eq_TC}) and the discussion below this equation). Thus, as the system approaches the superfluid instability, $-{\rm Re}\bigl[\Gamma^{\rm R}(\bm{q}=0,\nu=2\mu)\bigr]$ usually increases to diverge at $T_{\rm env}^{\rm c}$, as seen in Fig. \ref{fig9}(b).
\par
However, as one decreases $T_{\rm env}$ along the path (c) drawn in Fig. \ref{fig9}(a), Fig. \ref{fig9}(c) shows that $-{\rm Re}\bigl[\Gamma^{\rm R}(\bm{q},\nu=2\mu)\bigr]$ gradually has a peak at {\it non-zero} momentum ($\equiv {\bm q}_{\rm FF}$). This indicates the enhancement of FFLO-type pairing fluctuations characterized by ${\bm q}_{\rm FF}$ \cite{Kinninen2018}, although the present system has no spin imabalnce \cite{PhysRevA.73.051603,Chevy2010,PhysRevA.90.053629,Kinninen2018}. (We will explain the origin of this anomalous phenomenon in Sec.III.B.4.) In the equilibrium mean-field theory, the FFLO phase transition is known to occur, when $-{\rm Re}\bigl[\Gamma^{\rm R}(\bm{q}_{\rm FF}\ne 0,\nu=2\mu)\bigr]$ diverges \cite{Fulde1964,Larkin1964,Takada1969,Shimahara2007}. 
\par
However, as pointed out in Refs. \cite{Shimahara1998,Ohashi2002,Wang2018}, the FFLO phase transition does not actually occur in the present uniform system, because FFLO-pairing fluctuations destroy the long-range order. Thus, in Fig. \ref{fig9}(c), although the peak at ${\bm q}_{\rm FF}\ne 0$ continues to develop at low temperatures (which physically means the enhancement of fluctuations in the FFLO channel), it {\it never} diverges even at $T_{\rm env}=0$. For the lower superfluid phase transition temperature in the re-entrant region, with decreasing the non-equilibrium parameter $\delta\mu$ along the path (d) drawn in Fig. \ref{fig9}(a), the peak position moves to ${\bm q}=0$, as shown in Fig. \ref{fig9}(d). When the system reaches $T_{\rm env}^{\rm c}$, $-{\rm Re}\bigl[\Gamma^{\rm R}(\bm{q},\nu=2\mu)\bigr]$ diverges at ${\bm q}=0$. It means that this superfluid phase transition is, not the FFLO-type, but the ordinary BCS-type being characterized by the uniform superfluid order parameter $\Delta$. 
\par
Although FFLO-type pairing fluctuations prevent the system from the superfluid phase transition \cite{Shimahara1998,Ohashi2002,Wang2018}, they still contribute to the pseudogap phenomenon. Indeed, as shown in Fig. \ref{fig2}, the pseudogap regime between $T_{\rm env}^{\rm c}$ and $T_{\rm env}^*$ can be divided into the following two: 
\begin{enumerate}
\item[(a)] BCS-type pseudogap regime (PG$_{\rm BCS}$): The pseudogap is caused by BCS-type pairing fluctuations that are the strongest at ${\bm q}=0$. 
\item[(b)] FFLO-type pseudogap regime (PG$_{\rm FFLO}$): The pseudogap is caused by FFLO-type pairing fluctuations that are the strongest at ${\bm q}_{\rm FF}\ne0$. 
\end{enumerate}
Figure \ref{fig2}(a) shows that, when $(p_{\rm F}a_s)^{-1}=-0.6$, the low temperature pseudogap regime is dominated by FFLO-type pairing fluctuations. At the unitarity ($(p_{\rm F}a_s)^{-1}=0$), this regime shrinks (see panel (b)). The PG$_{\rm FFLO}$ region eventually disappears in the BEC regime, although we do not show the result here.
\par
Figure \ref{fig2}(a) consistently explains the phenomena (1) and (2) mentioned at the beginning of this subsection: The anomalous enhancement of FFLO-type pairing fluctuations around (c4) in Fig. \ref{fig6}(a) and (f4) in Fig. \ref{fig6}(d) gives large pseudogap seen in Figs. \ref{fig6}(b2) and (e2) at low temperatures, respectively. In addition, because of the pseudogap phenomenon associated with FFLO-type pairing fluctuations, $T_{\rm env}^*$ monotonically decreases with increasing $\delta\mu$ in the BCS regime where $T_{\rm env}^{\rm c}$ exhibits the re-entrant phenomenon.
\par
\begin{figure}[t]
\centering
\includegraphics[width=8cm]{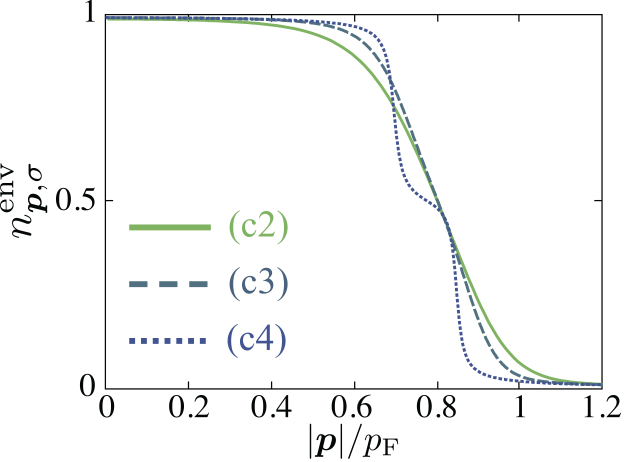} 
\caption{Momentum distribution $n_{{\bm p},\sigma}^{\rm env}$ of Fermi atoms, given in the first line in Eq. (\ref{eq.3}). We take $(p_{\rm F}a_s)^{-1}=-0.6$, and the results are at (c2)-(c4) in Fig. \ref{fig6}(a). In the case of (c4), $\mu/\ep_{\rm F}=0.602$ and $\delta\mu/\ep_{\rm F}=0.112$, so that the Fermi momenta of the pumping bath and decay bath equal $p_{\rm Fp}=\sqrt{2m[\mu+\delta\mu]}=0.847p_{\rm F}$ and $p_{\rm Fd}=\sqrt{2m[\mu-\delta\mu]}=0.697p_{\rm F}$, respectively.
}
\label{fig10} 
\end{figure}
\par
\subsubsection{Origin of FFLO-type pairing fluctuations}
\par
To explain the origin of FFLO-type pairing fluctuations enhanced in the non-equilibrium steady state, we show in Fig. \ref{fig10} the momentum distribution $n_{{\bm p},\sigma}^{\rm env}$ of Fermi atoms at (c2)-(c4) in Fig. \ref{fig6}(a). In Fig. \ref{fig10}, we find that $n_{{\bm p},\sigma}^{\rm env}$ has a two-step structure, when $T_{\rm env}=0$ (dotted line). This is a combined effect of the system-bath coupling with the vanishing thermal broadening of the momentum distribution at $T_{\rm env}=0$. Indeed, the Fermi momentum of the decay bath $p_{\rm Fd}=\sqrt{2m[\mu-\delta\mu]}=0.697p_{\rm F}$ and that of the pumping bath $p_{\rm Fp}=\sqrt{2m[\mu+\delta\mu]}=0.847p_{\rm F}$ are close to the momenta at the first and second (almost) vertical parts of $n_{{\bm p},\sigma}^{\rm env}$ at $T_{\rm env}=0$. 
\par
In a sense, this two-step structure is similar to the case of a spin-polarized Fermi gas, where two Fermi surfaces with different Fermi momenta $p_{\rm Fd}$ and $p_{\rm Fp}$. Of course, a crucial difference from the latter is that the present unpolarized case looks as if each spin component has two Fermi surfaces. However, as in the spin-polarized case, besides the Cooper pairs with zero center-of-mass-momentum, $|{\bm p}_{\rm Fd},\uparrow\rangle|-{\bm p}_{\rm Fd},\downarrow\rangle$ and $|{\bm p}_{\rm Fp},\uparrow\rangle|-{\bm p}_{\rm Fp},\downarrow\rangle$(this type of pairing is known as Sarma or breached-pair state \cite{Sarma1963,Liu2003} in the spin-polarized case), the FFLO-type pairs with nonzero center-of-mass-momentum, $|{\bm p}_{\rm Fd},\uparrow\rangle|-{\bm p}_{\rm Fp},\downarrow\rangle$ and $|{\bm p}_{\rm Fp},\uparrow\rangle|-{\bm p}_{\rm Fd},\downarrow\rangle$, become possible. Although the FFLO superfluid state is absent in the present uniform system \cite{Shimahara1998,Ohashi2002,Wang2018}, the two-step structure of the atomic momentum distribution may enhance fluctuations in the FFLO channel, giving PG$_{\rm FFLO}$ regime in Fig. \ref{fig2}.
\par
The two-step structure gradually becomes obscure with increasing $T_{\rm env}$ or decreasing $\delta\mu$. When pairing fluctuations become the strongest at ${\bm q}=0$, the BCS-type superfluid instability can occur. This also explains the re-entrant behavior of $T_{\rm env}^{\rm c}$ in the BCS-unitary regime.
\par
Since we are considering a population-balanced Fermi gas, the present mechanism of FFLO pairing is different from the conventional one discussed in superconductivity under an external magnetic field \cite{Fulde1964,Larkin1964,Takada1969,Shimahara2007}. Recently, the possibility of FFLO superfluid in the absence of external magnetic field has also been proposed, where the shift of the single-particle energy induced by external current \cite{PhysRevLett.97.257001}, or a size effect \cite{PhysRevLett.102.177001}, is used. The present case is also different from these previous proposals. Since the order parameter of the FFLO state has a spatially periodic pattern, it is rather close to a pattern formation discussed in other non-equilibrium systems \cite{PhysRevB.97.245302,Prigogine1977,RevModPhys.65.851,PhysRevLett.86.4163,PhysRevLett.100.250401,PhysRevX.7.031033}.
\par
\begin{figure}[tb]
\centering
\includegraphics[width=6.5cm]{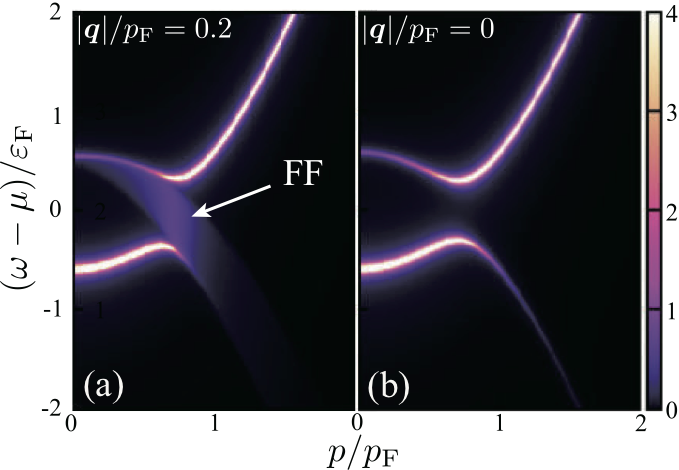}
\caption{Single-particle spectral weight $A({\bm p},\omega)$ (SW) in the static approximation. (a) Equation \eqref{Eq_static_FF} is used, assuming strong FFLO-type pairing fluctuations around $|\bm{q}_{\rm FF}|/p_{\rm F}=0.2$. A broad downward branch `FF' is consistent with Fig. \ref{fig6}(c4). (b) Equation \eqref{Eq_static} is used, assuming strong BCS-type pairing fluctuations around ${\bm q}=0$. In these model calculations, we set $\Delta_{\rm pg}/\ep_{\rm F}=0.3$, $\gamma/\ep_{\rm F}=0.01$, and use the same value of $\mu$ at (c4) in Fig. \ref{fig6}(a). The spectral intensity is normalized by $\ep_{\rm F}^{-1}$.
}
\label{fig11} 
\end{figure}
\par
\subsubsection{Spectral weight $A({\bm p},\omega)$ in PG$_{\rm FFLO}$ regime}
\par
In Sec. III.B.1, we used the static approximation, in order to simply explain the pseudogap phenomenon. To see how fluctuations in the FFLO channel affect single-particle excitations, we here extend this approximation to the case when pairing fluctuations are enhanced around ${\bm q}_{\rm FF}\ne 0$ and $\nu=2\mu$. In this case, the NETMA retarded self-energy is approximated to,
\begin{align}
\Sigma^\text{R}_\text{sys}(\bm{p},\omega) 
&\simeq 
-\Delta_{\rm pg}^2 
\langle G_{{\rm env},-\sigma}^{\rm A}({\bm q}_{\rm FF}-\bm{p},-\omega+2\mu)
\rangle_{{\bm q}_{\rm FF}}
\nonumber
\\
&=
-\frac{\Delta_\text{pg}^2}{4q_{\rm FF}p}
\log\left(\frac{\omega+(p-q_{\rm FF})^2
-\mu+2i\gamma}{\omega+(p+q_{\rm FF})^2-\mu+2i\gamma}\right),
\label{Eq_static_FF}
\end{align}
where $\langle\cdot\cdot\cdot\rangle_{{\bm q}_{\rm FF}}$ denotes the average over the direction of ${\bm q}_{\rm FF}$.
\par
Figure \ref{fig11}(a) shows SW in the case when the approximate self-energy in Eq. (\ref{Eq_static_FF}) (${\bm q}_{\rm FF}\ne 0$) is used. Comparing this result with the case when Eq. (\ref{Eq_static}) (${\bm q}=0$) is used (panel (b)), we see an additional downward broad spectrum `FF' in panel (a). The difference between the two originates from the fact that, while the hole branch $\omega-\mu=-[\varepsilon_{-{\bm p}}-\mu]$ only couples with the particle branch $\omega-\mu=[\varepsilon_{\bm p}-\mu]$ in the latter case, FFLO-type pairing fluctuations lead to the coupling between the particle branch $\omega-\mu=[\varepsilon_{\bm p}-\mu]$ and hole branches $\omega-\mu=-[\varepsilon_{{\bm q}_{\rm FF}-{\bm p}}-\mu]$ with {\it various directions} of ${\bm q}_{\rm FF}$. The spectral structure `FF' in Fig. \ref{fig11}(a) is thus regarded as a characteristic phenomenon associated with FFLO-type pairing fluctuations. 
\par
This additional spectral structure is indeed seen in the PG$_{\rm FFLO}$ regime, as shown in Fig. \ref{fig6}(c4). In the unitary limit, although the same structure should appear in Fig. \ref{fig6}(f4), it can not be clearly seen, due to the blurry SW by the strong pairing interaction.
\par
\begin{figure}[tb]
\centering
\includegraphics[width=6.5cm]{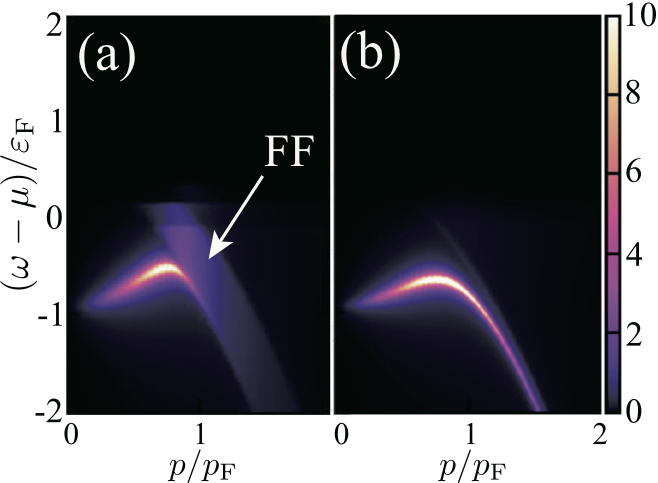}
\caption{(a) Calculated photoemission spectrum (PES) $L({\bm p},\omega)$ in a non-equilibrium Fermi gas, when $(p_{\rm F}a_s)^{-1}=-0.6$. (a) shows PES in the PG$_{\rm FFLO}$ regime at (c4) in Fig. \ref{fig6}(a). `FF' is related to the downward broad branch in Fig. \ref{fig6}(c4). (b) shows the result at the superfluid phase transition (c5) in Fig. \ref{fig6}(a), where pairing fluctuations are the strongest at ${\bm q}=0$. The spectral intensity is normalized by $[p_{\rm F}^2\ep_{\rm F}]^{-1}$. 
}
\label{fig12} 
\end{figure}
\par
The spectral structure `FF' in Fig. \ref{fig6}(c4) may be observed in the photoemission spectrum (PES), as shown in Fig. \ref{fig12}(a). For comparison, we also show in Fig. \ref{fig12}(b) PES corresponding to Fig. \ref{fig6}(c5). Because the latter is at $T_{\rm env}^{\rm c}$, pairing fluctuations are enhanced around ${\bm q}=0$. Because of this, the additional spectral structure is absent in this figure. Figure \ref{fig12} indicates that the photoemission-type expriment \cite{Stewart2008,PhysRevLett.114.075301,Torma2016} may be useful for the observation of this non-equilibrium phenomenon.
\par
\begin{figure}[tb]
\centering
\includegraphics[width=8.5cm]{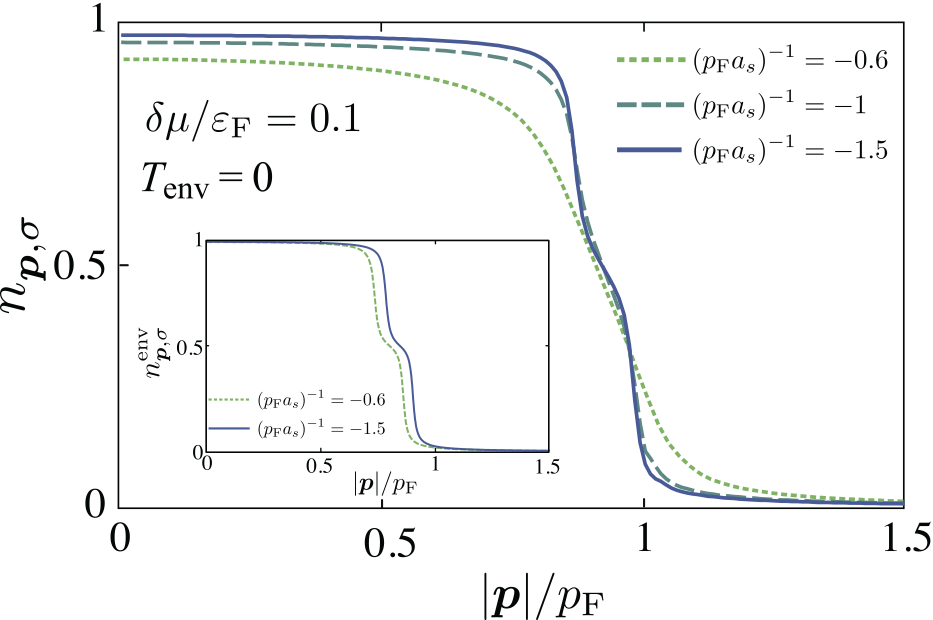}
\caption{Momentum distribution $n_{{\bm p},\sigma}$ of Fermi atoms, given in Eq. (\ref{eq.NETMA}). We set $\delta\mu/\varepsilon_{\rm F}=0.1$ and $T_{\rm env}=0$. The inset shows $n^{\rm env}_{{\bm p},\sigma}$, given in the first line in Eq. (\ref{eq.3}).
}
\label{fig13} 
\end{figure}
\par
\par
\begin{figure*}[btp]
\centering
\includegraphics[width=17cm]{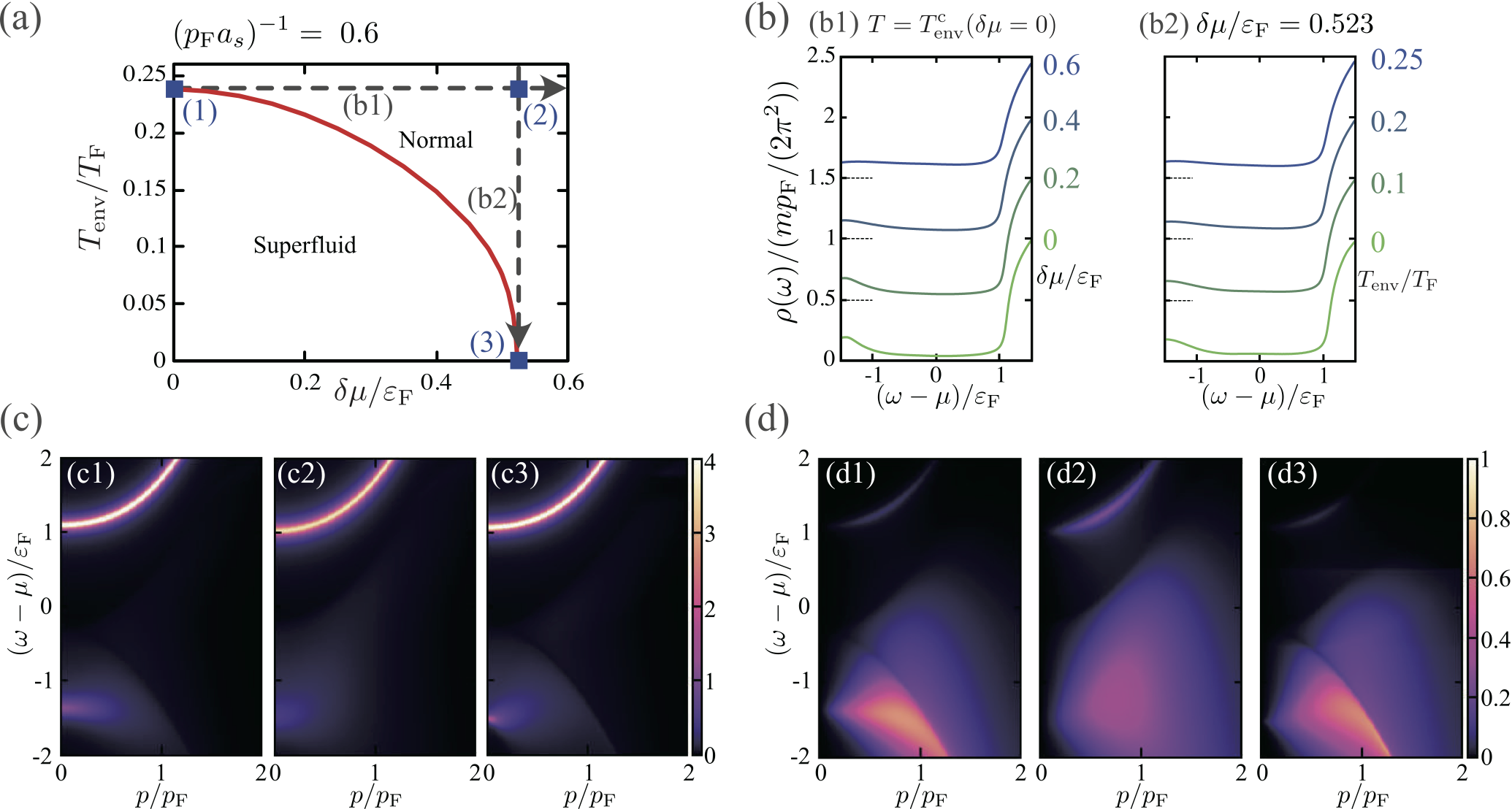}
\caption{(a) Calculated $T_{\rm env}^{\rm c}$ in the BEC regime ($(p_{\rm F}a_s)^{-1}=0.6$) where $T_{\rm env}^{\rm c}$ does not exhibit the re-entrant behavior.  (b) DOS $\rho(\omega)$. The left and right panels show the results along the paths (b1) and (b2) drawn in panel (a), respectively. In these panels, we offset the results. The short horizontal line near each result is at $\rho(\omega)=0$. (c) SW $A({\bm p},\omega)$. (d) PES $L(\bm{p},\omega)$. Panels (c$i$) and (d$i$) ($i$=1-3) show the results at the position $(i)$ in panel (a).
}
\label{fig14} 
\end{figure*}
\par
Before ending this subsection, we note that NETMA uses, not the fully dressed Green's function ${\hat G}_{{\rm sys},\sigma}$ in Eq. (\ref{Eq_G_sys}), but $\hat{G}_{{\rm env},\sigma}$ in Eq. (\ref{Eq_G_env}), in evaluating the diagrams in Fig. \ref{fig3}. To see influences of this, we calculate the momentum distribution $n_{{\bm p},\sigma}$ of Fermi atoms, from the fully dressed NETMA Green's function as
\begin{equation}
n_{{\bm p},\sigma}=-i\int_{-\infty}^\infty
{d\omega \over 2\pi}G_{{\rm sys},\sigma}^<({\bm p},\omega),
\label{eq.NETMA}
\end{equation}
where the NETMA lesser Green's function $G^<_{{\rm sys},\sigma}({\bm p},\omega)$ is given in Eq. (\ref{eq.G_less}). When $(p_{\rm F}a_s)^{-1}=-0.6$, Fig. \ref{fig13} shows that $n_{{\bm p},\sigma}$ does not exhibits the two-step structure, although $n_{{\bm p},\sigma}^{\rm env}$ does (see the inset of this figure). Since the two-step structure is crucial for the enhancement of FFLO-type pairing fluctuations, this result indicates that NETMA overestimates these fluctuations. However, Fig. \ref{fig13} also shows that this structure appears in $n_{{\bm p},\sigma}$, when $(p_{\rm F}a_s)^{-1}\lesssim -1$. Thus, we expect that the PG$_{\rm FFLO}$ regime would become narrower but still exist, even when we treat pairing fluctuations within the {\it self-consistent} $T$-matrix approximation (SCTMA) \cite{Haussmann1993,PhysRevB.61.15370}, where the fully dressed Green's function is used in the diagrams in Fig. \ref{fig3}. This improvement remains as our future challenge. 
\par
\subsection{BEC regime}
\par
We next consider the strong-coupling BEC regime where $T_{\rm env}^{\rm c}$ does not exhibit the re-entrant behavior. Figure \ref{fig14} shows an example of this situation (see panel (a)). In the BEC regime, system properties are dominated by tightly bound molecules, so that anomalous enhancement of FFLO-type pairing fluctuations associated with the detailed (two-step) structure of the atomic momentum distribution does not occur. As explained previously, in the BCS-unitary regime, the initial decrease of $T_{\rm env}^{\rm c}$ from the thermal equilibrium value $T_{\rm env}^{{\rm c}0}$ is due to the broadening of the atomic momentum distribution by the system-bath coupling, the effect of which is similar to the thermal effect in the thermal equilibrium case. The monotonic decrease of $T_{\rm env}^{\rm c}$ in Fig. \ref{fig14}(a) indicates that a similar effective thermal effect also works in the BEC regime, when $\delta\mu>0$.
\par
When $(p_{\rm F}a_s)^{-1}= 0.6$ shown in Fig. \ref{fig14}, DOS exhibits a large pseudogap in the thermal equilibrium state at $T_{\rm env}^{\rm c}$ (see Figs. \ref{fig14}(b1) and (c1)) \cite{PhysRevA.80.033613}. Figures \ref{fig14}(b) and (c) show that this pseudogap structure almost remains unchanged, as one moves along the paths (b1) and (b2). These results mean that the system is still dominated by tightly bound molecules even in the non-equilibrium state, at least in the parameter region in Fig. \ref{fig14}(a). 
\par
Figures \ref{fig14}(d1)-(d3) show PES at (1)-(3) in panel (a). Compared to panel (d1), the spectral intensity of the upper branch in panel (d2) is slightly enhanced. This is because the pumping bath at $T_{\rm env}$ supplies unpaired Fermi atoms to the system in the high-energy region $0\lesssim\omega-\mu\lesssim \delta\mu+T_{\rm env}$. Since our calculations fix the total number $N$ of atoms in the main system, this means that the molecular fraction is somehow smaller in the case of panel (d2) than panel (d1). Because the energy region where the pumping bath supplies atoms to the system becomes narrower at lower temperatures, the intensity of the upper branch again becomes weak in Fig. \ref{fig14}(d3). 
\par
To conclude, except for the above-mentioned suppression of $T_{\rm env}^{\rm c}$ and the enhancement of the upper particle branch in PES, non-equilibrium effects on single-particle excitations seem not so remarkable in the strong-coupling BEC regime as in the BCS-unitary case.
\par
\section{Summary}
\par
To summarize, we have discussed strong-coupling properties of a driven-dissipative two-component Fermi gas in the non-equilibrium steady state. We considered an open Fermi gas which is coupled with two baths: One supplies atoms to the system (pumping bath) and the other absorbs atoms from the system (decay bath). Non-equilibrium effects are tuned by adjusting the difference of the Fermi chemical potential between the two baths. In this model, we have also assumed a tunable $s$-wave pairing interaction associated with a Feshbach resonance.
\par
To theoretically treat both strong-pairing fluctuations and non-equilibrium effects, we have extended the $T$-matrix approximation (TMA), developed in  thermal equilibrium BCS-BEC crossover physics, to the non-equilibrium steady state. We have achieved this extension by employing the Keldysh Green's function theory.
\par
We showed that the superfluid phase transition temperature $T_{\rm env}^{\rm c}$ is suppressed when the system is out of equilibrium. While this suppression occurs in the whole BCS-BEC crossover region, we found that the detailed dependence of $T_{\rm env}^{\rm c}$ on the non-equilibrium parameter $\delta\mu$ (which equals half the difference of the chemical potential between the two baths) is very different between the BCS-unitary regime and the strong-coupling BEC regime. While $T_{\rm env}^{\rm c}$ monotonically decreases with increasing $\delta\mu$ in the latter regime, it exhibits re-entrant behavior in the former.\par
We pointed out that a two-step structure of the atomic momentum distribution, which is caused by the pumping and decay baths, is the origin of the re-entrant phenomenon: Although the present system has no spin polarization, this structure brings about a similar situation to a spin-polarized Fermi gas, where the Fermi-surface size of $\uparrow$-spin atoms is different from that of $\downarrow$-spin atoms. That is, the system becomes close to the FFLO instability. Although FFLO state is not actually stabilized in the present uniform case, this situation anomalously enhances fluctuations in the FFLO channel in the BCS-unitary regime, leading to the non-monotonic re-entrant behavior of $T_{\rm env}^{\rm c}$ as a function of $\delta\mu$. We also showed that, besides the ordinary BCS-type pairing fluctuations, FFLO-type pairing fluctuations also cause the pseudogap phenomenon. Since the enhancement of FFLO-type pairing fluctuations does not occur in the thermal equilibrium state of a spin-unpolarized Fermi gas, it is peculiar to the present non-equilibrium state. As a possible method to observe the pseudogap phenomenon associated with FFLO-type pairing fluctuations, we pointed out that the photoemission-type experiment may be useful.
\par
We end by listing some future problems. The non-equilibrium strong-coupling theory developed in this paper is only valid for the normal state. Extension to the superfluid phase below $T_{\rm env}^{\rm c}$ is an interesting future problem. Since the non-equilibrium steady state has recently been realized in a Fermi gas loaded on an moving optical lattice \cite{Chong2018}, extension of the present theory to the moving lattice system would also be important, in order to catch up this recent experimental progress. We also note that, because our theory is based on the {\it non-selfconsistent} $T$-matrix approximation (TMA), the fully dressed Green's function is not used in the TMA self-energy. To improve this, replacing TMA by the self-consistent $T$-matrix approximation (SCTMA) \cite{Haussmann1993,PhysRevB.61.15370} would be effective. Since the recent progress in cold Fermi gas experiments has enabled us to examine various interesting non-equilibrium phenomena, our results would contribute to the further development of cold atom physics in this direction.
\par
\par
\begin{acknowledgments}
We thank K. Manabe, R. Sato, and K. Furutani for discussions. RH was supported by a Grand-in-Aid for JSPS fellows (No.17J01238). YO was supported by a Grant-in-Aid for Scientific Research from MEXT and JSPS in Japan (No.JP18K11345, No.JP18H05406, and No.JP19K03689).
\end{acknowledgments}
\par
\appendix
\begin{widetext}
\section{Derivation of ${\hat \Sigma}_{\rm env}({\bm p},\omega)$ in Eq. (\ref{Eq_Sigma_env})}
\par
\par
In this appendix, we derive the $2\times 2$ matrix self-energy ${\hat \Sigma}_{\rm env}({\bm p},\omega)$ in Eq. (\ref{Eq_Sigma_env}), describing effects of the system-bath coupling.  Within the second-order Born approximation shown in Fig. \ref{fig3}(c), it is given by
\begin{equation}
{\hat \Sigma}_{{\rm env},\sigma}({\bm p},{\bm p}',t-t')
=
\sum_{\alpha={\rm L,R}}
|\Lambda_\alpha|^2
\sum_{i,j}^{N_{\rm t}}
\sum_{{\bm p},{\bm p'},{\bm q},{\bm q}'}
e^{i[{\bm q}\cdot{\bm R}_i^\alpha-{\bm p}\cdot{\bm r}^\alpha_i]}
{\hat D}_\sigma^\alpha({\bm q},{\bm q}',t-t')
e^{-i[{\bm q}\cdot{\bm R}_j^\alpha-{\bm p}\cdot{\bm r}^\alpha_j]}
e^{-i\mu_\alpha[t-t']},
\label{eq.a1}
\end{equation}
where ${\hat D}_\sigma^\alpha({\bm q},{\bm q}',t-t')$ is the $2\times 2$ matrix single-particle Keldysh Green's function in the $\alpha$-bath ($\alpha={\rm L,R}$). When we take the spatial average over the randomly distributing tunneling positions ${\bm R}_i^\alpha$ and ${\bm r}_i^\alpha$, the terms with $i=j$ only remain non-zero, and the translational invariance is recovered as $\langle {\hat \Sigma}_{{\rm env},\sigma}({\bm p},{\bm p'},t-t')\rangle_{\rm av}={\hat \Sigma}_{{\rm env},\sigma}({\bm p},t-t')\delta_{{\bm p},{\bm p}'}$. Here, 
\begin{equation}
{\hat \Sigma}_{{\rm env},\sigma}
({\bm p},t-t')=
N_{\rm t}\sum_\alpha|\Lambda_\alpha|^2
\sum_{\bm q}{\hat D}_\sigma^\alpha({\bm q},{\bm q},t-t')
e^{-i\mu_\alpha[t-t']}.
\label{eq.a2}
\end{equation}
\par
Carrying out the Fourier transformation with respect to $t-t'$, one has
\begin{eqnarray}
{\hat \Sigma}_{{\rm env},\sigma}({\bm p},\omega)
=
N_{\rm t}\sum_{\alpha={\rm L,R}}|\Lambda_\alpha|^2
\sum_{\bm q}{\hat D}_\sigma^\alpha({\bm q},\omega-\mu_\alpha),
\label{eq.aa1}
\end{eqnarray}
where ${\hat D}_\sigma^\alpha({\bm q},{\bm q}',\omega-\mu_\alpha)={\hat D}_\sigma^\alpha({\bm q},\omega-\mu_\alpha)\delta_{{\bm q},{\bm q}'}$, with
\begin{eqnarray}
{\hat D}_\sigma^\alpha({\bm q},\omega)=
\left(
\begin{array}{cc}
{1 \over \omega_+-\xi_{\bm q}^\alpha}
&
-2\pi i\delta(\omega-\xi_{\bm q}^\alpha)
\tanh\left({\omega \over 2T_{\rm env}}\right)
\\
0
&
{1 \over \omega_--\mu_\alpha-\xi_{\bm q}^\alpha}
\end{array}
\right).
\label{eq.DD}
\end{eqnarray}
We thus obtain
\begin{eqnarray}
{\hat \Sigma}_{{\rm env},\sigma}({\bm p},\omega)
=
N_{\rm t}\sum_{\alpha={\rm L,R}}|\Lambda_\alpha|^2
\sum_{\bm q}
\left(
\begin{array}{cc}
{1 \over \omega_+-\mu_\alpha-\xi_{\bm q}^\alpha}
&
-2\pi i\delta(\omega-\mu_\alpha-\xi_{\bm q}^\alpha)
\tanh\left({\omega-\mu_\alpha \over 2T_{\rm env}}\right)
\\
0
&
{1 \over \omega_--\mu_\alpha-\xi_{\bm q}^\alpha}
\end{array}
\right).
\label{eq.a3}
\end{eqnarray}
Assuming the white baths with the constant density of states $\rho$, as well as replacing the ${\bm q}$-summation in Eq. (\ref{eq.a3}) by the $\xi^\alpha$-integration, we obtain Eq. (\ref{Eq_Sigma_env}).
\par
\par
\section{Derivation of NETMA self-energies $\Sigma_{\rm NETMA}^{\rm R,K,A}({\bm p},\omega)$ in Eqs. (\ref{Eq_Sigma_sys_R}) and (\ref{Eq_Sigma_sys_K})} 
\par
In this appendix, we explain how to obtain the NETMA self-energies $\Sigma_{\rm NETMA}^{\rm R,K,A}({\bm p},\omega)$ in Eqs. (\ref{Eq_Sigma_sys_R}) and (\ref{Eq_Sigma_sys_K}). 
\par
\subsection{Pair correlation functions $\Pi^{\rm R,K,A}({\bm q},\nu)$ in Eqs. (\ref{Eq_Pi_R}) and (\ref{Eq_Pi_K})}
\par
For later convenience, we first derive Eqs. (\ref{Eq_Pi_R}) and (\ref{Eq_Pi_K}). For this purpose, we introduce the following $2\times 2$ matrix pair correlation function:
\begin{eqnarray}
\left(
\begin{array}{cc}
\Pi^{--}({\bm q},\nu) &\Pi^{-+}({\bm q},\nu)\\
\Pi^{+-}({\bm q},\nu) &\Pi^{++}({\bm q},\nu)\\
\end{array}
\right)
=
-i
\int {d\nu \over 2\pi}e^{i\nu [t-t']}
\left(
\begin{array}{cc}
\langle
{\hat T}_{\rm t}
\{\Delta_{\bm q}(t)\Delta^\dagger_{-{\bm q}}(t')
\}
\rangle 
&
\langle
\Delta^\dagger_{-{\bm q}}(t')\Delta_{\bm q}(t)
\rangle 
\\
\langle
\Delta_{\bm q}(t)\Delta^\dagger_{-{\bm q}}(t')
\rangle &
\langle
{\hat T}_{\rm at}
\{\Delta_{\bm q}(t)\Delta^\dagger_{-{\bm q}}(t')
\}
\rangle 
\end{array}
\right),
\nonumber
\\
\label{eq.b1}
\end{eqnarray}
where ${\hat T}_{\rm t}$ (${\hat T}_{\rm at}$) represent the time-ordered (anti-time-ordered) product, and
\begin{equation}
\Delta_{\bm q}=
\sum_{\bm p}
c_{{-{\bm p}+{\bm q}/2},\downarrow}
c_{{{\bm p}+{\bm q}/2},\uparrow}.
\label{eq.b2}
\end{equation}
In the zero-th order with respect to the pairing interaction $-U$, Eq. (\ref{eq.b1}) can be written as,
\begin{eqnarray}
\left(
\begin{array}{cc}
\Pi^{--}({\bm q},\nu) &\Pi^{-+}({\bm q},\nu)\\
\Pi^{+-}({\bm q},\nu) &\Pi^{++}({\bm q},\nu)\\
\end{array}
\right)
=
i
\sum_{\bm p}
\int {d\omega \over 2\pi}
\left(
\begin{array}{cc}
G^{--}_{{\rm env},\downarrow}(p_+)G^{--}_{{\rm env},\uparrow}(p_-)
&
G^{-+}_{{\rm env},\downarrow}(p_+)G^{-+}_{{\rm env},\uparrow}(p_-)
\\
G^{+-}_{{\rm env},\downarrow}(p_+)G^{+-}_{{\rm env},\uparrow}(p_-) 
&
G^{++}_{{\rm env},\downarrow}(p_+)G^{++}_{{\rm env},\uparrow}(p_-)
\end{array}
\right).
\nonumber
\\
\label{eq.b3}
\end{eqnarray}
Here, we have introduced the abbreviated notation, $p_\pm=(\pm{\bm p}+{\bm q}/2,\pm\omega+\nu/2)$. The single-particle Green's functions $G_{{\rm env},\sigma}^{\pm\pm}$ are related to $G_{{\rm env},\sigma}^{\rm R,K,A}$ in Eq. (\ref{Eq_G_env}) through the Keldysh rotation as \cite{Rammer2007,Zagoskin,Stefanucci},
\begin{eqnarray}
\left(
\begin{array}{cc}
G^{\rm R}_{{\rm env},\sigma} & G^{\rm K}_{{\rm env},\sigma} \\
0 & G^{\rm K}_{{\rm env},\sigma}
\end{array}
\right)
=
{\hat K}
\tau_3
\left(
\begin{array}{cc}
G^{--}_{{\rm env},\sigma} & G^{-+}_{{\rm env},\sigma} \\
G^{+-}_{{\rm env},\sigma} & G^{++}_{{\rm env},\sigma}
\end{array}
\right)
{\hat K}^\dagger,
\label{eq.b4}
\end{eqnarray}
where ${\hat K}=[1-i\tau_2]/\sqrt{2}$ with $\tau_i$ being the Pauli matrix. The correlation functions $\Pi^{\pm\pm}$ are also related to $\Pi^{\rm R,K,A}$ through the same Keldysh rotation. Using these relations ($\Pi^{\pm\pm}\leftrightarrow\Pi^{\rm R,K,A}$ and $G^{\pm\pm}_{{\rm env},\sigma}\leftrightarrow G^{\rm R,K,A}_{{\rm env},\sigma}$), we obtain Eqs. (\ref{Eq_Pi_R}) and (\ref{Eq_Pi_K}). In this procedure, we have used the following analytic property:
\begin{equation}
\int_{-\infty}^\infty d\omega
G^{\rm R}_{{\rm env},\sigma}(p_+)
G^{\rm A}_{{\rm env},-\sigma}(p_-)=0.
\label{eq.b4b} 
\end{equation}
\par
\begin{figure}[t]
\centering
\includegraphics[width=13cm]{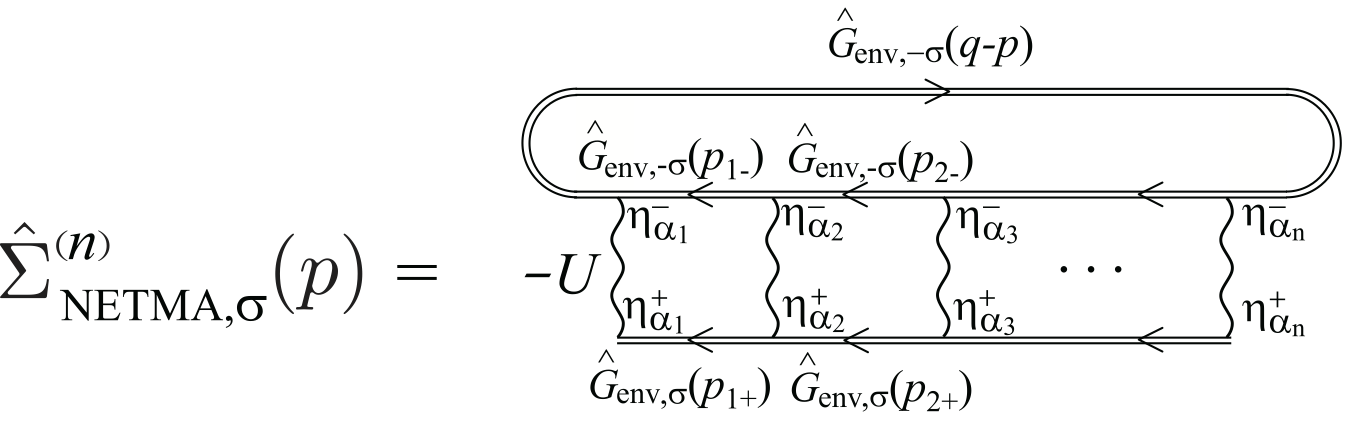} 
\caption{Diagrammatic structure of NETMA self-energy $\hat{\Sigma}^{(n)}_{\rm NETMA}({\bm p},\omega)$ in the last term in Fig. \ref{fig3}(a). $p_{i\pm}=(\pm {\bm p}_i+{\bm q}/2,\pm\omega_i+\nu/2)$, and $\eta^\pm_{\alpha=1,2}$ is defined in Eq. (\ref{eq.b6}).}
\label{fig15} 
\end{figure}
\par
\subsection{NETMA self-energies in Eqs. (\ref{Eq_Sigma_sys_R}) and (\ref{Eq_Sigma_sys_K})} 
\par
We next derive the NETMA self-energy in Eqs. (\ref{Eq_Sigma_sys_R}) and (\ref{Eq_Sigma_sys_K}). Figure \ref{fig15} shows the $n$-th order self-energy diagram ${\hat \Sigma}^{(n)}_{{\rm NETMA},\sigma}({\bm p},\omega)$ with respect to the pairing interaction $-U$, which gives, 
\begin{align}
\hat{\Sigma}^{(n)}_{{\rm NETMA},\sigma}(p) 
&=-
(-iU)^n\int dq
\sum_{{\alpha_1^\pm,\alpha_2^\pm,\cdot\cdot\cdot,\alpha_n^\pm=1,2}}
\int \prod_{i=1}^{n-1} dp_i
\eta_{\alpha_1}^+
{\hat G}_{{\rm env},\sigma}(p_{1+})
\eta_{\alpha_2}^+
{\hat G}_{{\rm env},\sigma}(p_{2+})
\cdot\cdot\cdot
\eta_{\alpha_n}^+
\nonumber
\\
&\times
{\rm Tr}
\left[
\eta_{\alpha_1}^-
{\hat G}_{{\rm env},-\sigma}(p_{1-})
\eta_{\alpha_2}^-
{\hat G}_{{\rm env},-\sigma}(p_{2-})
\cdot\cdot\cdot
\eta_{\alpha_n}^-
{\hat G}_{{\rm env},-\sigma}(q-p)
\right].
\label{eq.b5}
\end{align}
Here, we have used the abbreviated notations, $p=({\bm p},\omega)$, $q=({\bm q},\nu)$, $p_{i\pm}=({\bm p}_i+{\bm q}/2,\pm\omega_i+\nu/2)$, $\int dq=\sum_{\bm q}\int d\nu/(2\pi)$, and $\int dp_i=\sum_{{\bm p}_i}\int d\omega_i/(2\pi)$. The vertex matrices $\eta_{\alpha=1,2}^\pm$ are given by \cite{Rammer2007}
\begin{eqnarray}
\left.
\begin{array}{l}
\eta_\alpha^+={1 \over \sqrt{2}}\tau_{2-\alpha}, \\
\eta_\alpha^-={1 \over \sqrt{2}}\tau_{\alpha-1}.
\end{array}
\right.
\label{eq.b6}
\end{eqnarray}
\par
We rewrite Eq. (\ref{eq.b5}) in the form,
\begin{eqnarray}
\left[\Sigma_{{\rm NETMA},\sigma}^{(n)}(p)\right]_{a,b}
&=&
-{i \over 2}
\sum_{c,d}\int dq
\Gamma^{(n)}_{(a,b),(c,d)}(q)
G_{{\rm env},-\sigma}^{d,c}(q-p)
\nonumber
\\
&=&
-{i \over 2}
\int dq
{\rm Tr}
\left[{\tilde \Gamma}_{a,b}^{(n)}(q)
{\hat G}_{{\rm env},-\sigma}(q-p)
\right],
\label{eq.b7}
\end{eqnarray}
where
\begin{eqnarray}
{\tilde \Gamma}^{(n)}_{a,b}(q)
&=&
\left(
\begin{array}{cc}
\Gamma^{(n)}_{(a,b),(1,1)}(q) & \Gamma^{(n)}_{(a,b),(1,2)}(q) \\
\Gamma^{(n)}_{(a,b),(2,1)}(q) & \Gamma^{(n)}_{(a,b),(2,2)}(q)
\end{array}
\right)
\nonumber
\\
&=&
\left(
\begin{array}{cccc}
\Gamma^{(n)}_{(11),(11)}(q) & 
\Gamma^{(n)}_{(11),(12)}(q) &
\Gamma^{(n)}_{(12),(11)}(q) & 
\Gamma^{(n)}_{(12),(12)}(q) \\
\Gamma^{(n)}_{(11),(21)}(q) & 
\Gamma^{(n)}_{(11),(21)}(q) &
\Gamma^{(n)}_{(12),(21)}(q) & 
\Gamma^{(n)}_{(12),(22)}(q) \\
\Gamma^{(n)}_{(21),(11)}(q) & 
\Gamma^{(n)}_{(21),(12)}(q) &
\Gamma^{(n)}_{(22),(11)}(q) & 
\Gamma^{(n)}_{(22),(12)}(q) \\
\Gamma^{(n)}_{(21),(21)}(q) & 
\Gamma^{(n)}_{(21),(22)}(q) &
\Gamma^{(n)}_{(22),(21)}(q) & 
\Gamma^{(n)}_{(22),(22)}(q) \\
\end{array}
\right).
\label{eq.b8}
\end{eqnarray}
Each matrix element in this equation can be written by using the Kronecker product $\otimes$ as,
\begin{eqnarray}
\Gamma_{(a,b),(c,d)}^{(n)}(q)
&=&
-2i(-iU)^n
\sum_{{\alpha_1^\pm,\alpha_2^\pm,\cdot\cdot\cdot,\alpha_n^\pm=1,2}}
\int \prod_{i=1}^{n-1} dp_i
\left[
\eta_{\alpha_1}^+
{\hat G}_{{\rm env},\sigma}(p_{1+})
\eta_{\alpha_2}^+
{\hat G}_{{\rm env},\sigma}(p_{2+})
\cdot\cdot\cdot
\eta_{\alpha_n}^+
\right]_{a,b}
\nonumber
\\
&\times&
\left[
\eta_{\alpha_1}^-
{\hat G}_{{\rm env},-\sigma}(p_{1-})
\eta_{\alpha_2}^-
{\hat G}_{{\rm env},-\sigma}(p_{2-})
\cdot\cdot\cdot
\eta_{\alpha_n}^-
{\hat G}_{{\rm env},-\sigma}(q-p)
\right]_{c,d}
\nonumber
\\
&=&
-2i(-iU)^n
\sum_{{\alpha_1^\pm,\alpha_2^\pm,\cdot\cdot\cdot,\alpha_n^\pm=1,2}}
\int \prod_{i=1}^{n-1} dp_i
\Bigl[
\left[
\eta_{\alpha_1}^+
{\hat G}_{{\rm env},\sigma}(p_{1+})
\eta_{\alpha_2}^+
{\hat G}_{{\rm env},\sigma}(p_{2+})
\cdot\cdot\cdot
\eta_{\alpha_n}^+
\right]
\nonumber
\\
&\otimes&
\left[
\eta_{\alpha_1}^-
{\hat G}_{{\rm env},-\sigma}(p_{1-})
\eta_{\alpha_2}^-
{\hat G}_{{\rm env},-\sigma}(p_{2-})
\cdot\cdot\cdot
\eta_{\alpha_n}^-
{\hat G}_{{\rm env},-\sigma}(q-p)
\right]
\Bigr]_{(a,b),(c,d)}.
\label{eq.b9}
\end{eqnarray}
Using the formula,
\begin{equation}
({\hat A}{\hat C})\otimes ({\hat B}{\hat D}) =
({\hat A}\otimes {\hat B})
({\hat C} \otimes {\hat D}),
\label{eq.b10}
\end{equation}
we further rewrite Eq. (\ref{eq.b9}) as,
\begin{eqnarray}
\Gamma^{(n)}_{(a,b),(c,d)}(q)
&=&
(-U)^n
\left[
{\hat \eta}
[{\tilde \Pi}(q){\hat \eta}]^n
\right]_{(a,b),(c,d)}.
\label{eq.b11}
\end{eqnarray}
Here,
\begin{equation}
{\hat \eta}=2\sum_{\alpha=1,2}
\eta_\alpha^+\otimes\eta_\alpha^-
=
\left(
\begin{array}{cc}
\tau_1 &\tau_0 \\
\tau_0 &\tau_1
\end{array}
\right),
\label{eq.b12}
\end{equation}
and
\begin{eqnarray}
{\tilde \Pi}(q)
&=&
{i \over 2}
\int dp
{\hat G}_{{\rm env},\sigma}(p_+)
\otimes
{\hat G}_{{\rm env},-\sigma}(p_-)
\nonumber
\\
&=&
\left(
\begin{array}{cccc}
{\tilde \Pi}_{(11),(11)}(q) & 
{\tilde \Pi}_{(11),(12)}(q) &
{\tilde \Pi}_{(12),(11)}(q) & 
{\tilde \Pi}_{(12),(12)}(q) \\
0 & 
0 &
0 & 
{\tilde \Pi}_{(12),(22)}(q) \\
0 & 
0 &
0 & 
{\tilde \Pi}_{(22),(12)}(q) \\
0 & 
0 &
0 & 
{\tilde \Pi}_{(22),(22)}(q) \\
\end{array}
\right).
\label{eq.b13}
\end{eqnarray}
The vanishing matrix elements in Eq. (\ref{eq.b13}) are the consequences of the absence of the (2,1) component of the Keldysh Green's function ${\hat G}_{{\rm env},\sigma}$ in Eq. (\ref{Eq_G_env}), as well as the the analytic property in Eq. (\ref{eq.b4b}). Comparing the detailed expressions for the non-vanishing components in Eq. (\ref{eq.b13}) with $\Pi^{\rm R,K,A}(q)$ in Eqs. (\ref{Eq_Pi_R}) and (\ref{Eq_Pi_K}), we obtain the following relations between them:
\begin{align}
\Pi^{\rm R}(q)&={\tilde \Pi}_{(11),(12)}(q)+{\tilde \Pi}_{(12),(11)}(q),
\label{Eq_Pi_R_app}\\
\Pi^{\rm A}(q)&={\tilde \Pi}_{(12),(22)}(q)+{\tilde \Pi}_{(22),(12)}(q),
\label{Eq_Pi_A_app}\\
\Pi^{\rm K}(q)&={\tilde \Pi}_{(11),(11)}(q)+{\tilde \Pi}_{(12),(12)}(q)
+{\tilde \Pi}_{(22),(22)}(q).
\label{Eq_Pi_K_app}
\end{align}
\par
The self-energy ${\hat \Sigma}_{{\rm NETMA},\sigma}(p)$ is then obtained from the summation of Eq. (\ref{eq.b8}), which gives
\begin{equation}
{\hat \Sigma}_{{\rm NETMA},\sigma}(p)=
\sum_{n=1}^\infty {\hat \Sigma}^{(n)}_{{\rm NETMA},\sigma}(p)
=-{i \over 2}\int dq 
{\rm Tr}\left[{\tilde \Gamma}(q){\hat G}_{{\rm env},-\sigma}(q-p)\right],
\label{eq.b14}
\end{equation}
where
\begin{eqnarray}
{\tilde \Gamma}(q)
=\sum_{n=1}^\infty{\tilde \Gamma}^{(n)}(q)
&=&-{\hat \eta}
{U \over 1+U{\tilde \Pi}(q){\hat \eta}}
=
\left(
\begin{array}{cccc}
0 & \Gamma^{\rm A}(q) & \Gamma^{\rm A}(q) & 0 \\
\Gamma^{\rm R}(q) & \Gamma^{\rm K}(q) & 
\Gamma^{\rm K}(q) & \Gamma^{\rm R}(q) \\
\Gamma^{\rm R}(q) & \Gamma^{\rm K}(q) & 
\Gamma^{\rm K}(q) & \Gamma^{\rm R}(q) \\
 0 & \Gamma^{\rm A}(q) & \Gamma^{\rm A}(q) & 0 
\end{array}
\right).
\label{eq.b15}
\end{eqnarray}
In obtaining the last expression, we have used the relations in Eqs. (\ref{Eq_Pi_R_app})-(\ref{Eq_Pi_K_app}), as well as the expressions for $\Gamma^{\rm R,K,A}(q)$ in Eq. (\ref{Eq_Gamma}). 
\par
The NETMA self-energies $\Sigma_{{\rm NETMA},\sigma}^{\rm R,K,A}(p)$ are then obtained from the (1,1), (1,2), and (2,2) components of Eq. (\ref{eq.b14}), respectively. For example, the retarded component $\Sigma_{{\rm NETMA},\sigma}^{\rm R}(p)=[{\hat \Sigma}_{{\rm NETMA},\sigma}^{\rm R}(p)]_{11}$ is given by
\begin{align}
\Sigma^{\rm R}_{{\rm NETMA},\sigma}(p) 
&= -\frac{i}{2} \sum_{\bm{q}} \int_{-\infty}^\infty \frac{d\nu}{2\pi}  
{\rm Tr}\Bigl[ \tilde{\Gamma}_{11}(q) \hat{G}_{{\rm env},-\sigma}(-p+q)\Bigr]
\nonumber\\
&= -\frac{i}{2} \sum_{\bm{q}} \int_{-\infty}^\infty \frac{d\nu}{2\pi} {\rm Tr}
\left[
\left(
\begin{array}{cc}
0 & \Gamma^{\rm A}(q) \\
\Gamma^{\rm R}(q) & \Gamma^{\rm K}(q)
\end{array}
\right)
\left(
\begin{array}{cc}
G_{{\rm env},-\sigma}^{\rm R}(-p+q) & G_{{\rm env},-\sigma}^{\rm K}(-p+q) \\
0 & G_{{\rm env},-\sigma}^{\rm A}(-p+q)
\end{array}
\right)
\right] 
\nonumber\\
&=-\frac{i}{2} \sum_{\bm{q}} \int_{-\infty}^\infty \frac{d\nu}{2\pi} 
\Bigl[ \Gamma^{\rm R}(q) G_{{\rm env},-\sigma}^{\rm K}(-p+q) + \Gamma^{\rm K}(q) G_{{\rm env},-\sigma}^{\rm A}(-p+q)  \Bigr].
\end{align}
Equations (\ref{Eq_Sigma_sys_R}) and (\ref{Eq_Sigma_sys_K}) can also be obtained in the same manner. We briefly note that $[\hat{\Sigma}_{{\rm NETMA},\sigma}(p)]_{21}$ vanishes, as expected.
\end{widetext}
\par

\end{document}